\documentclass[acmsmall]{acmart}
\usepackage{tikz}
\usepackage{amsmath} 
\usepackage{graphicx}
\usepackage{caption}
\usepackage{subcaption}
\usepackage{tabularx}
\usepackage{makecell}
\usepackage{hyperref}
\usepackage{csquotes}

\usepackage{xcolor}

\newcommand{\dquote}[1]{{``\emph{#1}''}}

\setcopyright{cc}
\setcctype[4.0]{by-nc-sa}
\acmJournal{PACMHCI}
\acmYear{2025} \acmVolume{9} \acmNumber{7} \acmArticle{CSCW498} \acmMonth{11} \acmDOI{10.1145/3757679}

\begin{document}

% \title{Mixed Reality Street Play: Towards a Playful City with Ubiquitous Mixed Reality Head-Mounted Display}

% \title[Towards Immersive Mixed Reality Street Play]{Towards Immersive Mixed Reality Street Play: Understanding Co-located Bodily Play with See-through Head-Mounted Displays in Public Spaces}

\title{Towards Immersive Mixed Reality Street Play}
\subtitle{Understanding Co-located Bodily Play with See-through Head-mounted Displays in Public Spaces}

%New title suggestion: "Mixed Reality Street Play: Mapping of the Design Space of Multiplayer Mixed Reality Games in the Wild with Head-mounted Displays"

% Synchronous Asymmetric

%% "author note" and "authornotemark" commands

\author{Botao Amber Hu}\authornote{Corresponding author}
\orcid{0000-0002-4504-0941}
\affiliation{%
  \institution{Reality Design Lab}
  \city{New York City}
  \country{USA}
  }
\email{botao@reality.design}

\author{Rem RunGu Lin}
\orcid{0000-0003-1931-7609}
\affiliation{%
  \institution{The Hong Kong University of Science and Technology (Guangzhou)}
  \city{Guangzhou}
  \country{China}
  }
\email{rlin408@connect.hkust-gz.edu.cn}

\author{Yilan Elan Tao}
\orcid{0000-0003-1691-9727}
\affiliation{%
  \institution{Reality Design Lab}
  \city{New York City}
  \country{USA}
  }
\email{elan@reality.design}

\author{Samuli Laato}
\orcid{0000-0003-4285-0073}
\affiliation{%
  \institution{University of Turku}
  \city{Turku}
  \country{Finland}
  }
\email{sadala@utu.fi}

\author{Yue Li}
\orcid{0000-0003-3728-218X}
\affiliation{%
  \institution{Xi’an Jiaotong-Liverpool University}
  \city{Suzhou}
  \country{China}
  }
\email{yue.li@xjtlu.edu.cn}

% \renewcommand{\shortauthors}{Botao Amber Hu et al}

\begin{abstract}
%% abstract < = 150 words
%% abstract < = 150 words
%% abstract < = 150 words

As see-through \emph{Mixed Reality Head-Mounted Displays} (MRHMDs) proliferate, their usage is gradually shifting from controlled, private settings to spontaneous, public contexts. While location-based augmented reality mobile games such as \emph{Pokémon GO} have been successful, the embodied interaction afforded by MRHMDs moves play beyond phone-based screen-tapping toward co-located, bodily, movement-based play. In anticipation of widespread MRHMD adoption, major technology companies have teased concept videos envisioning urban streets as vast mixed reality playgrounds---imagine Harry Potter--style wizard duels in city streets---which we term \emph{Immersive Mixed Reality Street Play} (IMRSP). However, few real-world studies examine such scenarios. Through empirical, in-the-wild studies of our research-through-design game probe, \emph{Multiplayer Omnipresent Fighting Arena} (MOFA), deployed across diverse public venues, we offer initial insights into the social implications, challenges, opportunities, and design recommendations of IMRSP. The MOFA framework, which includes three gameplay modes---``The Training'', ``The Duel'', and ``The Dragon''---is open-sourced at \url{https://github.com/realitydeslab/mofa}.
\end{abstract}

\begin{CCSXML}
<ccs2012>
<concept>
  <concept_id>10003120.10003123.10011758</concept_id>
  <concept_desc>Human-centered computing~Interaction design theory, concepts and paradigms</concept_desc>
  <concept_significance>100</concept_significance>
</concept>
<concept>
  <concept_id>10003120.10003121.10003124.10010392</concept_id>
  <concept_desc>Human-centered computing~Mixed / augmented reality</concept_desc>
  <concept_significance>500</concept_significance>
</concept>
<concept>
<concept_id>10003120.10003130.10003233</concept_id>
<concept_desc>Human-centered computing~Collaborative and social computing systems and tools</concept_desc>
<concept_significance>100</concept_significance>
</concept>
</ccs2012>
\end{CCSXML}

\ccsdesc[100]{Human-centered computing~Interaction design theory, concepts and paradigms}
\ccsdesc[500]{Human-centered computing~Mixed / augmented reality}
\ccsdesc[100]{Human-centered computing~Collaborative and social computing systems and tools}

\keywords{Social Augmented Reality, Collaborative Mixed Reality, Co-located Mixed Reality, Bystander Inclusion, Social Acceptance, Game Design, Urban Games, Pervasive Games, Co-located Bodily Play, See-through Head-mounted Display, Research in the Wild, Public Spaces, Spontaneous Encounters, Awkwardness, Safety Concerns}

\received{October 2024}
\received[revised]{April 2025}
\received[accepted]{August 2025}

\begin{teaserfigure}
    \centering
    \includegraphics[width=1\linewidth]{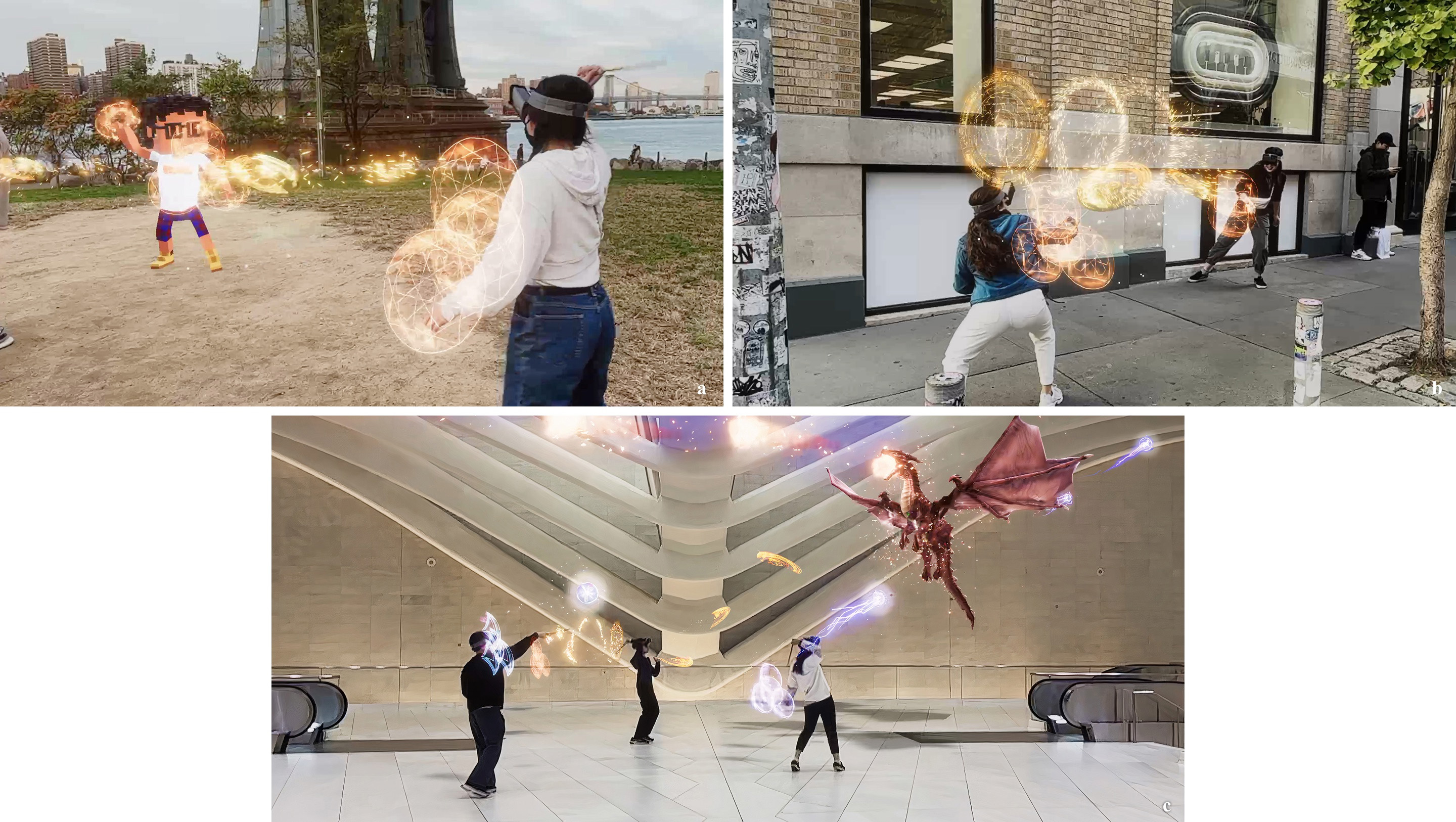}
    \caption{Multiplayer Omnipresent Fighting Arena (MOFA) is a research-through-design game probe series studying the social implications of Immersive Mixed Reality Street Play (IMRSP)---the scenario of co-located bodily play with mixed reality see-through head-mounted displays in public spaces. We deployed MOFA in the wild across three public venues: (a) "The Training" in a public park in Dumbo, Brooklyn; (b) "The Duel" on the urban streets of SoHo, Manhattan; and (c) "The Dragon" in the public atrium of World Trade Center Transportation Hub, New York City. }
    \label{fig:gameplay}
\end{teaserfigure}

\maketitle

\section{Introduction}
\label{sec:introduction}

Major technology companies such as Apple and Meta are rapidly advancing see-through \emph{Mixed Reality Head-Mounted Display} (MRHMD) technology: Apple's Vision Pro \cite{apple2023visionpro} offers video see-through mixed reality in a headset form factor, while Meta's Orion \cite{meta2024orion} boasts optical see-through capabilities in an eyeglasses‑style form factor. Meta CEO Mark Zuckerberg has even predicted that \emph{``smart glasses will replace phones by 2030''} \cite{birch2024smartglasses}. Some scholars forecast that augmented reality could become virtually indistinguishable from physical reality in certain contexts~\cite{Nijholt2023Everpresent,rauschnabel2021}, underscoring the importance of understanding user experiences and social dynamics surrounding these emerging devices. As Itoh et al.~\cite{Itoh2021Indistinguishable} demonstrate, passthrough technologies---whether video see-through or optical see-through---enable users to maintain awareness of their real-world surroundings while moving freely, extending potential usage from controlled private settings to spontaneous public scenarios. This shift raises pressing questions about how ubiquitous MRHMD adoption could transform everyday public life in the near future. 

% In the current augmented reality landscape, Pokémon GO stands as the most successful application, generating \$6 billion over eight years~\cite{iqbal2024pokemon}. While Pokémon GO's promotional videos~\footnote{``Discover Pokémon in the Real World with Pokémon GO!
% '', \url{https://www.youtube.com/watch?v=2sj2iQyBTQs}, visited on October 25, 2024} depict a future where people engage in device-free bodily play on public streets, the game in today's market still primarily relies on traditional smartphone screen-touching interactions. Moreover, some scientific studies estimate that less than 10\% of active Pokémon GO players use the game's AR features~\cite{laato2021}. Despite this, Niantic, the company behind Pokémon GO, continues to release technical demo videos~\footnote{For one such video, see ``Lightship x Snapdragon Spaces | Niantic'' \url{https://www.youtube.com/watch?v=sDNjVxX0LrE}, visited on October 25, 2024} showcasing a future where people engage in MRHMD-based bodily play on public streets, interacting directly with nearby friends~\footnote{``Niantic | Meet You Out There'' \url{https://www.youtube.com/watch?v=HljcLVXAxAU}, visited on October 25, 2024}. We term this vision \emph{``Immersive Mixed Reality Street Play''} (IMRSP)—a form of co-located bodily play in public spaces using MRHMDs that primarily involves body movements as the main mode of interaction. Although the technology to implement such games exists, there's a significant research gap regarding empirical studies for MRHMD-based bodily play in the wild. Therefore, in this study, we explore the feasibility of this vision through the following research question (RQ):

Games are widely anticipated to be one of the ``killer apps'' of this envisioned MR future \cite{Rhodes2019}. The most successful augmented reality title to date is \emph{Pokémon GO}, which has generated over \$6 billion in eight years~\cite{iqbal2024pokemon}.
While Pokémon GO's promotional videos\footnote{``Discover Pokémon in the Real World with Pokémon GO!'', \url{https://www.youtube.com/watch?v=2sj2iQyBTQs}, visited on October 25, 2024} depict a future where people engage in device-free bodily play on public streets, games in today's market still primarily rely on traditional smartphone screen-touching interactions. Moreover, some scientific studies estimate that less than 10\% of active Pokémon GO players use the game's AR features~\cite{Laato2021Convergence}. Despite this, Niantic, the game’s developer, continues to release speculative promotional videos\footnote{``Niantic | Meet You Out There'', \url{https://www.youtube.com/watch?v=HljcLVXAxAU}, visited on October 25, 2024} and tech demo videos\footnote{ ``Lightship x Snapdragon Spaces | Niantic,'' \url{https://www.youtube.com/watch?v=sDNjVxX0LrE}, visited on October 25, 2024}
that depict a future in which people seamlessly engage in bodily gameplay using see-through MR devices---or even device-free interfaces---spontaneously interacting with friends on city streets. We refer to this scenario as \emph{Immersive Mixed Reality Street Play} (IMRSP), a form of co-located bodily play in public spaces using MRHMDs that primarily involves body movements as the main mode of interaction. IMRSP is not a new concept and has been popularized in mainstream media, including commercial apps such as \emph{Pokémon GO} and science fiction films such as \emph{Free Guy}. Although the MR technology to implement IMRSP exists, real-world deployments---particularly MRHMD-based bodily play involving spontaneous participants in urban environments---remain underexplored. Therefore, in this study, we explore this vision through the following research question (RQ):

\begin{quote}
    \emph{What social implications might arise when immersive mixed reality street play becomes prevalent in public spaces in a future with ubiquitous see‑through head‑mounted displays?}
\end{quote}

We developed \emph{Multiplayer Omnipresent Fighting Arena} (MOFA)\footnote{Fun fact: The pronunciation of ``MOFA'' coincidentally means ``magic'' in Chinese.} as a research-through-design  \cite{Zimmerman2007Research} game probe to investigate IMRSP in real-world contexts. MOFA is an open-source\footnote{MOFA is open-sourced at \url{https://github.com/realitydeslab/mofa}.} MR game framework that goes beyond traditional location-based, screen-touching mobile gaming by enabling co-located, multiplayer bodily play through MRHMDs. To better support in-the-wild studies, we introduced MOFA's ``omnipresent'' feature: unlike other co-located MR games that require local Wi-Fi, this feature enables gameplay to occur spontaneously anywhere---even without Wi-Fi, cellular, or Internet connectivity---by using Apple's Multipeer Connectivity technology, as detailed in Section \ref{sec:system}. The framework centers on spell-casting gestures with a physical wand. This magic-themed interaction serves three key purposes: First, the spell-casting gestures become intuitive once users recognize the Harry Potter--inspired context, even if the movements initially seem unusual; Second, ranged attacks using virtual magical energy balls avoid physical contact, improving safety in urban settings during our in-the-wild deployments; Third, the magical theme enables various social interactions---from solo practice to multiplayer duels---and creates a cohesive narrative across all three MOFA experiences through consistent spell-casting interactions: ``\emph{The Training}'' (single-player mode for practicing spell-casting skills), ``\emph{The Duel}'' (two-player wizard dueling), and ``\emph{The Dragon}'' (three or more players cooperatively hunting a virtual monster). 

We adopted an in-the-wild approach  \cite{Benford2013Performance} to maximize ecological validity \cite{Chaytor2003Ecological}, deploying MOFA across diverse real-world public venues---such as university campuses, tech expos \cite{Hu2024MOFA, Hu2023MOFA}, urban parks, public building courtyards, and natural wilderness sites---and inviting passersby to participate. We used opportunistic recruitment in these public spaces, recording the spectator-view AR feed of each gameplay session and conducting brief post-play contextual inquiries with both players and onlookers to capture immediate reactions. 
Additionally, to gain broader perspectives beyond those of on-site participants, we performed a desktop review of online forum discussions generated after sharing our actual gameplay videos. This allowed us to compare firsthand in-the-wild experiences with the more detached insights of remote desktop observers, ultimately examining whether the IMRSP vision, initially sparked by those speculative promotional videos, truly holds up in real-world settings.

Unlike traditional mobile-based AR games, IMRSP requires substantial bodily movement in public spaces. While this enables presence and collaborative excitement, it also introduces distinct ethical, social acceptability, privacy, and safety concerns~\cite{Millard2024Ethics, Katins2024Assessing}. The online review feedback revealed more concerns than those expressed by on-site participants. As IMRSP games are physically demanding and often challenge established social norms---for example, due to the power imbalance between MRHMD wearers and bystanders~\cite{Chung2023Negotiating}---the long-term player retention of IMRSP in public spaces remains uncertain. These implications underscore that the path to the ubiquitous IMRSP vision may involve friction unless we address these issues head-on. 
Our empirical studies provide an initial yet foundational understanding of IMRSP. By examining how MRHMD technologies reshape social norms, dynamics, interactions, and collaborative play in public spaces, our findings inform both future research directions and design guidelines for creating more responsible and inclusive public IMRSP experiences.

In summary, our key contributions are: 

\begin{itemize}
\item An initial real-world exploration of social dynamics and phenomena in IMRSP, revealing key benefits and challenges of this novel form of play;
\item A comparative analysis of online reviews versus on-site gameplay experiences, highlighting the friction of social acceptability between idealized visions and practical reality;
\item Design recommendations and a research agenda for future IMRSP development derived from our exploratory work;
\item An open-source framework MOFA for spontaneous co-located multiplayer MR gameplay that facilitates further IMRSP research.
\end{itemize}
 % amber
\section{Background and Related Work}
\label{sec:background}

% key message: 
% 1. Street play 
% 各种类型的 Play in the Street, 
% Non-tech to Tech
% Urban play, playful City, Pervasive play \cite{HinskeClassifying}, 
% 

% Street play
% 1. 各种类型 from non tech to tech 
% 2.   Urban play, playful city, human - city interaction. 
% 3. LBS Game: Pokémon go, geocaching. 
% 4. Can you see me now? Mixed Reality. 
% 5. Funny and natural setting and movement. 
% 6. Bystander inclusion problem.
% 7. Social acceptance problem. 
% 8. Safety and health concerns 
% 9. Spontaneous encountering problem . Human robotics. 

% Bodily play
% 1. Button-pressing button to co-located bodily play . 
% 2. It’s movement based games. Muller. Somatic turn. 
% 3. Co-located bodily play. 
% 4. Funny. Social dynamics and synergy. Health. Exertion games . 
% 5. Find happiness and improve social interaction. 
% 6. 

% MR HMD in public spaces 
% 1. MR HMD and see through. 
% 2. Imbalanced power 
% Soocisl concerns :
% Awkward: unscripted 
% Social acceptance. 
% Healthy issue
% Privacy issues. 

% MR co-located and collaborated work 
% Asymmetrical 
% Bystander solving by phone spectating . 
% Issues how to get co-located technological challenges. 
% MR HMD bodily play
% 1. Social XR Weird Interaction. 
% 2. Astrie
% 3. Hado 

% Button-pressing games to co-located Bodily play 

\subsection{Technology-mediated Street Play}

Street play has long been an integral part of urban culture, transforming streets into dynamic spaces where social interactions, creativity, and spontaneity flourish \cite{tranter1996reclaiming}. Traditional street play, characterized by informal, unstructured activities like hopscotch, tag, or ball games \cite{wridt2004historical}, utilizes public spaces in ways that challenge the boundaries of social and spatial order within urban environments \cite{Stevens2007Ludic}. 
Streets become arenas for social interaction, where norms are negotiated and social bonds are strengthened \cite{Whyte1980Social}. The spontaneous nature of street play fosters inclusivity and accessibility, enabling people of various ages and backgrounds to participate and interact. Meanwhile, the fields of urban play \cite{Chisik2022Editorial}, playable city \cite{Nijholt2017Towards}, pervasive play \cite{Ahn2016Pervasive}, alternate reality games \cite{Kim2008Alternate}, and pervasive games \cite{Montola2009Pervasive} have all contributed to redefining human-city interactions ~\cite{Lee2021Augmented}.

Technological advances in network coverage, power consumption, and device miniaturization have paved the way for new forms of outdoor interactive computing \cite{Balestrini2020Moving}. These breakthroughs have spawned diverse deployments and studies of technology use in public spaces. Examples include large interactive displays \cite{Cao2008Flashlight, Hara2008Understanding}, mobile-based mixed reality \cite{Flintham2003Where}, location-based gaming (such as ``Can You See Me Now?'' \cite{Benford2006Can}, geocaching \cite{Ohara2008Geocaching}, and Pokémon GO \cite{Paavilainen2017Pokemon}), and human-robot interactions on streets \cite{Pelikan2024Encountering}. These technological interventions have transformed street play, introducing innovative forms of engagement that blend physical and digital realms seamlessly \cite{De2009Hybrid}.

The street is a site of socially organized human actions. Studies of behavior in public spaces often draw from Erving Goffman's work, which identifies social behaviors and the management of impressions in everyday interactions \cite{goffman2023presentation,goffman2008behavior}. Ethnomethodological and conversation analytic studies have examined public settings, emphasizing that social order is fragile and easily disrupted \cite{mcgrane1994tv}. 
Shusterman’s ``Bodies in the Streets'' \cite{Shusterman2019} examines how bodily sensations, practices, and perceptions profoundly shape our everyday experiences in urban environments, illuminating the somaesthetic dimensions of city life. These studies highlight how individuals utilize subtle bodily cues in street interactions to approach others, regardless of familiarity \cite{DeStefani2018Encounters}. Furthermore, visual information plays a crucial role in how people navigate the street, coordinating movements and avoiding collisions \cite{Hester2003Analysing}. 
Together, these insights illustrate the delicate nature of social interactions in urban environments and the embodied practices that facilitate social order amidst potential disruptions.

Informed by such studies, HCI has developed conceptual frameworks for thinking about designing for interaction in public, including research in the wild \cite{Chamberlain2012Research}, performance-led research \cite{Benford2013Performance}, and design considerations of bystanders and spectators in public interactions \cite{Reeves2005Designing,Reeves2011Designing}. These approaches have been instrumental in understanding and shaping user experiences in outdoor settings.
Meanwhile, HCI research has documented numerous social phenomena that arise when people interact with technology in public contexts, such as public displays~\cite{Hara2008Understanding} and public installations~\cite{Balestrini2016Jokebox}. Among these are the honeypot effect~\cite{Wouters2016Uncovering}, where an initial user attracts onlookers and encourages them to engage; showmanship and performance, where users become aware of an audience and amplify their actions ~\cite{Benford2011Performing,Peltonen2008Mine,Reeves2005Designing}; social mirroring, in which newcomers imitate observed behaviors~\cite{Peltonen2008Mine}. Researchers have also noted social discomfort~\cite{Benford2012Uncomfortable}, social facilitation (improved performance in front of others)~\cite{Peltonen2008Mine,Miller2019Social}, social awkwardness~\cite{Stefanidi2025Youre,Deterding2018Alibis,Huggard2013Musical,Press2023Humorous}, and inhibition (anxiety or reluctance to participate publicly)~\cite{Miller2019Social}.

\subsection{Technology-supported Co-located Bodily Play}
In recent years, HCI research has undergone a ``somatic turn'' \cite{Loke2018Somatic}. This shift places the human body at the center of interactive technology design, moving beyond traditional user-centered approaches \cite{Dix2017Human} to focus on the body as the primary site of interaction and experience \cite{Hook2018Designing}. Merleau-Ponty argues that the human body is fundamental to perception and action \cite{Merleau2008World}, with proprioception and kinesthesia serving as core movement-related senses \cite{Zhou2021Dance}. Homewood et al. \cite{Homewood2021Tracing} outline three evolving conceptions of the body in HCI: the individual body (embodied interaction), multiple bodies (plurality of experiences), and more-than-human bodies (entangled assemblages). These concepts, particularly entangled assemblages \cite{Forlano2017Posthumanism}, intercorporeality \cite{Meyer2017Intercorporeality,Stepanova2024Intercorporeal,Patibanda2024Exploring}, and post-phenomenology \cite{Ron2018Bowl}, provide theoretical frameworks for exploring technology-supported co-located bodily play.

The conceptual frameworks of proxemics \cite{Ciolek1983proxemics} and f-formations \cite{Barua2021Detecting} significantly inform the practical exploration of co-located bodily play. They offer insights into how spatial and social arrangements enhance interaction within shared physical spaces. These principles are instrumental in designing co-located bodily experiences that are socially engaging and spatially harmonious. Research on co-located bodily play \cite{Segura2015Enabling} in exertion games \cite{Kaos2019Socialb} further underscores the importance of multiplayer experiences and social interaction in enhancing game adherence and enjoyment. Co-located bodily play design is rooted in social bodily play \cite{Mueller20182nda, Mueller20192nd}, encompassing three key areas: bodily play \cite{Mueller2014Proxemics, Jorgensen2024Body, Martin-Niedecken2018Designing, Mueller2020Designing, MarquezSegura2013design, Isbister2018Social}, movement-based design \cite{Mueller2014Movementbased, vanRheden2024Why, VanDelden2023Technology, Isbister2015Guidelines, Garner2013Combining, Hamalainen2015Utilizing}, and social exertion games \cite{Mueller2017Designing, Sinclair2007Considerations, MarquezSegura2021Exploring, Mueller2017Five, Elvitigala2024Granda}. This body of work collectively emphasizes the integration of physical activity, social interaction, and immersive digital content in creating engaging and harmonious social play experiences.

\subsection{Co-located Mixed Reality}

% For early stage, there is early play to early ARQuake \cite{Piekarski2002ARQuake} explored outdoor augmented reality systems but relied on a heavy backpack setup that limited its ecological validity for real-world, in-the-wild testing. 

The focus of MR research has increasingly shifted from single-user experiences to multi-user collaborations, particularly within the CSCW community. Billinghurst and Kato coined the term ``Collaborative Mixed Reality'' to describe this emerging field \cite{Billinghurst2002Collaborativea}. Collaborative MR can be broadly categorized into remote collaboration and face-to-face co-located collaboration. 
Remote collaboration has been extensively studied \cite{Schafer2023Survey}. These studies have demonstrated MR's potential to bridge physical distances and enhance communication in distributed teams. On the other hand, face-to-face co-located collaboration in MR \cite{Billinghurst2002Experiments} has been explored in two main areas: interactions between MRHMD wearers \cite{Nilsson2009colocateda} and collaborative work involving both handheld devices and MRHMDs \cite{Zaman2023MRMACa}. The latter has particularly focused on bystander inclusion, addressing the challenge of integrating non-HMD users into MR experiences \cite{Zhu2020BISHARE, Denning2014situ}. Research on co-located collaboration between HMD wearers has investigated shared workspace designs, collaborative problem-solving, and social dynamics in MR environments \cite{Sereno2020Collaborative}. Auda et al. \cite{Auda2023Scoping} conducted a comprehensive scoping survey on these cross-reality systems.

Unlike traditional mobile games that rely on screen interactions---such as location-based multiplayer experiences like Pokémon GO \cite{Xu2023Understanding, Paavilainen2017Pokemon, Alha2019Why, Poretski2021Who} or co-located handheld AR games \cite{Yusof2019Collaborativea,Reig2023Supporting, Xu2008BragFish,Nijholt2021Experiencing}---MRHMD wearers naturally use their bodies as controllers. This shift from screen-touching to movement-based co-located bodily play represents a significant change in technology-mediated gameplay. While extensive research has explored co-located MR interactions \cite{Bhattacharyya2019Brick,Sereno2020Collaborative,Dagan2022Projecta,Lin2024Cell,Miedema2019Superhuman,Grandi2019Characterizinga,Zhou2019Astaire,Radu2021Survey,Thomas2012survey,He2019Exploring,DSouza2018Augmentingc,Reig2023Supporting}, these studies have primarily been conducted in controlled, private settings. Those studies highlight the need for designs that foster social inter-bodily engagement \cite{Mueller2018Experiencinga}. Notably, commercial location-based entertainment like HADO \cite{araki2018hado}, a multiplayer co-located dodgeball-like mixed reality sport, and Spatial Ops \cite{Spatialops}, marketed as the world's first competitive multiplayer shooter that \emph{``turns your home into a battlefield''}, have emerged. However, these games require specific setups and predefined environments---such as Wi-Fi connectivity or special marker walls---which limit their potential for spontaneous play in public or outdoor settings.
Despite some early academic explorations~\cite{Bonfert2017Invaders, Piekarski2002ARQuake}, to our knowledge, few commercially available or open-source systems currently explore co-located bodily play using MRHMDs in public spaces. 

\subsection{Mixed Reality Concerns in Public Spaces}
\label{subsection:concerns}

In recent years, MRHMDs with passthrough functionality~\cite{Itoh2021Indistinguishable} have sparked a notable shift from private, controlled settings to more public, spontaneous environments. These devices, including video and optical see-through technologies such as Apple Vision Pro, Magic Leap, Microsoft HoloLens, and Meta Orion, enable users to view their physical surroundings while interacting with virtual elements~\cite{Radu2021Survey,Bailenson2024Seeing}. 
The rise in MRHMD usage in public settings raises social concerns for both wearers and non-wearers~\cite{Gugenheimer2017ShareVR,Sapkota2021Ubiquitous}. 
The social acceptance~\cite{Koelle2020Social} of MRHMD usage in public spaces tends to decline significantly~\cite{Koutromanos2023Augmented,Profita2016Effect,Sun2025Experiencing,Schwind2018Virtual, Eghbali2019Social, Gugenheimer2019Challenges,Gugenheimer2022Novel,OHagan2023Augmenting,Lebeck2018Security} due to three main factors:

\paragraph{Imbalanced power dynamics} MRHMDs grant wearers capabilities---such as data overlays or video recording---that non-wearers cannot see or fully understand. This imbalance can negatively affect social experiences, privacy, and safety~\cite{Chung2023Negotiating}. Denning et al.~\cite{Denning2014situ} examined bystander privacy in AR, while O'Hagan et al.~\cite{OHagan2021Safety,Hagan2023Privacy} studied bystanders' privacy concerns and awareness regarding AR glasses. A notable example of these social tensions is Google Glass, which faced criticism for prioritizing user capabilities over broader social impacts. This led to the term ``Glasshole'', illustrating bystanders' perception of antisocial behavior by Google Glass wearers~\cite{Due2015social}.

\paragraph{Embodied interactions} MRHMDs often rely on embodied interactions~\cite{Handosa2018Extending}, such as mid-air gestures and full-body movements. Although these interactions can be intuitive for manipulating virtual objects, they may also contravene established social norms, engendering social awkwardness~\cite{Plakias2024Awkwardness,Krauss2024What,Chen2024Awkward,Guo2024Exploring}. Moreover, these conspicuous gestures~\cite{Montero2010Would,Ahlstrom2014Are} may exacerbate the power imbalance between wearers and bystanders, as onlookers may feel excluded or uncertain about the wearer's actions in mixed reality.

\paragraph{Social acceptance and safety in public spaces} When MRHMDs are used ``on the go'' in public spaces \cite{Schwind2018Virtual}, such as urban streets~\cite{Guo2024Exploring}, public libraries~\cite{Kaeder2024Working}, public transit~\cite{Bajorunaite2025Enacting,Bajorunaite2023Reality,Medeiros2022Shielding,Bajorunaite2024VR,Medeiros2024Social}, and airplanes~\cite{Ng2021Passenger,Williamson2019PlaneVR}, social acceptance, trust, security, and privacy become even more pressing \cite{Katins2024Assessing, Millard2024Ethics,Roesner2014Security,Zhu2024Make,Abraham2022Implications}. Interactions and movement in public traffic environments can pose additional risks, especially when MRHMD wearers navigate between physical and virtual elements. These interactions can intensify existing power imbalances and increase the potential for harms and social friction with bystanders.

% \added{
% \cite{Medeiros2023Surveying}
% \cite{OHagan2022PrivacyEnhancing}
% The Societal Harms Posed by Everyday Augmented Reality
% \cite{OHagan2024Viewpoint}
% \cite{Weerasinghe2025Mute}

%   title = {A {{Viewpoint}} on the {{Societal Impact}} of {{Everyday Augmented Reality}} and the {{Need}} for {{Perceptual Human Rights}}},
% }

\subsection{Research Gap: Immersive Mixed Reality Street Play}

While the concept of MR street play has been advocated for several years, there remains a significant gap in research regarding co-located bodily play with MRHMDs in public spaces. To our knowledge, no commercial MRHMD games currently exist that can be reliably deployed or played on the streets. Although extensive empirical research has examined the social implications of smartphone-based street play like Pokémon GO \cite{Alha2019Why}, our literature review reveals no structured understanding or empirical studies of IMRSP. This under-exploration appears to stem from three main reasons:

\paragraph{Inaccessibility of affordable MRHMDs}
\label{reason:inaccessibility}
As of 2025, MRHMDs with either full-color video passthrough or optical see-through---such as Vision Pro, HoloLens, and Meta's Orion---remain prohibitively expensive, with prices starting at \$3,500 per device. This high cost poses a significant barrier to co-located multiplayer play, which requires multiple MRHMDs. Consequently, research in this area is typically confined to controlled lab settings, limiting the exploration of in-the-wild applications.

\paragraph{Limited technology for spontaneous, low‑latency multiplayer mixed reality co-location}
\label{reason:spontaneous}
Multiplayer mixed reality co-location requires low-latency spatial pose transmission---less than 20 ms latency---to prevent noticeable lag \cite{Struye2020Latency}. This necessitates either deploying a Wi-Fi router in the wild or utilizing low-latency cellular networks like 5G, which do not yet provide ubiquitous coverage. Furthermore, routing data through remote servers between players often results in latency that exceeds user tolerance thresholds.

\paragraph{Social acceptance barriers and lack of established social norms} \label{reason:acceptance}
The current form factor of MRHMDs is far from resembling everyday glasses. Additionally, interaction behaviors such as mid-air gestures and full-body movements lack established social norms for use in public settings. This absence of social norms leads to social awkwardness and acceptance issues. Furthermore, commercial companies have little immediate market incentive to research this area.

\subsection{Methods for Exploring Future Mixed Reality}
This paper interrogates the prospective social implications of IMRSP as a form of play in public spaces in a future where MRHMDs are as ubiquitous as today's smartphones. Currently, no commercially viable IMRSP experiences can be deployed on real streets with ecological validity. Therefore, we turn to design futuring \cite{Fry2018Design} methods---approaches intended to provoke reflection on possible sociotechnical futures, reveal tensions and frictions \cite{Pierce2021Tension}, challenge assumptions, and stimulate critical dialogue \cite{Celik2025Reviewingc}.

Traditional design futuring~\cite{Fry2018Design,Kozubaev2020Expanding,Celik2025Reviewingc,Lindley2015Back} practices, including speculative design~\cite{Auger2013Speculativea,Dunne2013Speculativea,Barendregt2021Speculative}, design fiction~\cite{Sterling2009COVER, Blythe2014Researcha}, experiential futures~\cite{Candy2017Designing}, critical design~\cite{Bardzell2012Critical}, material speculation~\cite{Wakkary2015Material}, typically employ fictional or \emph{low‑fidelity} artifacts~\cite{Zhu2024Exploring,Gaver2012Whatb}, such as diegetic prototypes~\cite{Kirby2010Futurea}, mock‑ups~\cite{Blythe2016AntiSolutionist}, scripted scenarios~\cite{Candy2017Designing},  games~\cite{Coulton2016Games,Brown2016IKEA}, or concept videos~\cite{Briggs2012Invisible,Odom2024Negotiating,Sturdee2016Design} to provoke reflection, to imagine alternative worlds, and to foreground the discursive power. Such formats, however, often leave audiences on the far side of what Candy calls an ``experiential gulf'' \cite{Candy2010Futures}: viewers discuss a hypothetical future rather than live it, producing debate rather than situated behavioral data \cite{RingfortFelner2025Quality}. Elsden et al.~\cite{Elsden2017Speculativea} further caution that lab‑based user enactments~\cite{Odom2012fieldworka} can break the suspension of disbelief. To address this gap and facilitate experiential participation, \emph{Speculative Enactments}~\cite{Elsden2017Speculativea} stage scenarios as lived experiences that foreground real‑world consequences. Building on this, \emph{Immersive Speculative Enactments}~\cite{Simeone2022Immersive} leverage virtual reality technologies to place participants in open‑ended situations that invite authentic action rather than following fixed linear narratives, as is common in \emph{Immersive Design Fiction}~\cite{McVeighSchultz2018Immersivea}. Cools et al. \cite{Cools2024Impacta} offer an Immersive Speculative Enactment example in which participants experience fictional MR contact lens scenarios through a multi-user VR simulation, generating experiential insights into the technology’s potential impact on daily life.

Building on this lineage, we treat the future not as a scenario to be imagined but as a situation to be enacted and studied \emph{in situ}. Unlike many design futuring approaches, IMRSP is not a speculative impossibility; it is technically achievable today. We see opportunities to directly implement IMRSP through current capabilities of hardware and software systems. What limits our ability to observe its social implications is not feasibility but adoption: MRHMDs are not yet ubiquitous in everyday public life. We could implement \emph{high‑fidelity}, fully functional IMRSP prototypes to bring the future into the present and use them as research probes in public spaces, allowing us to study its real-world implications directly. 

This performative, embodied approach \cite{Buchenau2000Experience,Benford2013Performance} turned city streets into staged yet plausible near‑future environments in which passersby could actively participate. Through \emph{in‑the‑wild} trials \cite{Rogers2017Research}, we documented real‑time social behaviors, safety negotiations, and bystander--player dynamics---insights that fictional formats cannot surface. Our study thus transforms a \emph{what‑if} into a lived experiment: a form of speculative enactment in which participants \dquote{act on and experience the speculation, beyond commenting on it} \cite{Elsden2017Speculativea}. Conducted as Research through Design (RtD) \cite{Zimmerman2007Research}, the field deployment yields ecologically valid, embodied behavioral data and shows how combining speculative futures with rigorous \emph{in situ} experimentation can both reveal and shape trajectories for emerging technologies.

\section{System Design}
\label{sec:system}

To explore the real-world social implications of IMRSP, we employed a research-through-design approach. We developed a series of game prototypes called ``Multiplayer Omnipresent Fighting Arena'' (MOFA) and deployed these games in real-world scenarios. MOFA serves as an open-source MR game framework that enables co-located, multiplayer bodily play through MRHMDs, allowing spontaneous play on the street via peer-to-peer local wireless networking.

\subsection{Design Considerations}
To design gameplay probes that can actually run in the wild with current technology, we considered the following design factors to ensure the experiment's viability:

\subsubsection{Easy to Understand and Learn}
The game probe should be easily understood by city dwellers on the streets. Inspired by the \emph{Harry Potter} fantasy, particularly wizard dueling, we adopted a `magic' theme for the gameplay. This aligns well with the characteristics of augmented reality media---unseen energy invisible to ``Muggles'', who are unaware of magic. This theme draws from popular culture that has been deeply rooted in global consciousness for over three decades. It makes the rules and settings intuitively understandable for many without explanation in ``in-the-wild'' settings. Even when behaviors might seem unusual in public, they are explainable and comprehensible because people have been familiarized through Harry Potter movies—they recognize these wand dueling interactions. Players can instantly immerse themselves in imaginative wizard and witch roles.

\subsubsection{Physical Props for Social Affordance}
All gameplay for MRHMD wearers in MOFA incorporates a physical prop: a wand. We discovered that the wand provides an intuitive affordance for players to understand the hand-waving gesture used to cast spells. This proves effective for three primary reasons: (1) It aligns with the stereotypical image of wizardry popularized by the Harry Potter series, enabling people to easily grasp the interaction without tutorials; (2) The physical shape and weight of the wand facilitate a natural ``wave down'' gesture, which is more intuitive than any bare-handed gesture for casting spells (which typically require tutorials). With wands, the interaction becomes second nature; (3) The wands help passersby readily comprehend the context of the games, even without seeing the augmented reality imaging overlay. Addressing the social acceptance concerns described in Section \ref{subsection:concerns}, the wand serves as a strong social affordance~\cite{Isbister2018Social} for both players and passersby in public spaces.

\subsubsection{Co-located Bodily Play without Physical Contact}
As envisioned in promotional videos, MRHMD wearers engage in embodied interactions involving full-body movements, transcending traditional screen-touching gameplay. MOFA exemplifies co-located bodily play with MRHMDs, showcasing interactions ranging from single-player to multiplayer scenarios. The game deliberately avoids physical contact for two key reasons: (1) many existing mixed reality systems cannot detect physical contact between players, making tangible interactions like melee attacks impractical in game design; (2) to prioritize player safety, we designed the game with ranged attacks only, eliminating the risk of direct physical contact between players. MOFA strategically incorporates a magic shield in front of players to indicate an attacking target, which allows opponents to aim at the shield, providing a clear target without risking physical harm. Unlike traditional fighting games, where avatars display damage through fainting, blood spatter, or falling, MOFA uses a shattering shield visual effect to provide feedback when an opponent successfully lands a hit. This design ensures both safety and clear interaction feedback.

\subsubsection{Omnipresent Co-location}
MOFA is designed for deployment anywhere, transforming any street into a game arena---hence its name, ``omnipresent''. It should operate without Wi-Fi or cellular connections. Mixed reality co-location requires low-latency spatial pose transmission (under 20 ms) to prevent noticeable lag. To address the research gap in spontaneous co-location technology outlined in Section \ref{reason:spontaneous}, we employed an open-source co-location toolkit \cite{hu2023InstantCopresence}. This toolkit utilizes Multipeer Connectivity\footnote{\url{https://developer.apple.com/documentation/multipeerconnectivity}, accessed on October 25, 2024}, the technology powering Apple's AirDrop, to enable local networking. This innovation allows street play to begin anywhere, even in rural areas lacking cellular signals or Wi-Fi access. The technology also features a rapid QR scanning process for synchronizing coordinates across devices, eliminating the need to pre-scan the physical environment and enabling spontaneous player as well as spectator participation.

\subsection{Gameplay Overview}

\begin{figure}
    \centering
    \includegraphics[width=0.96\linewidth]{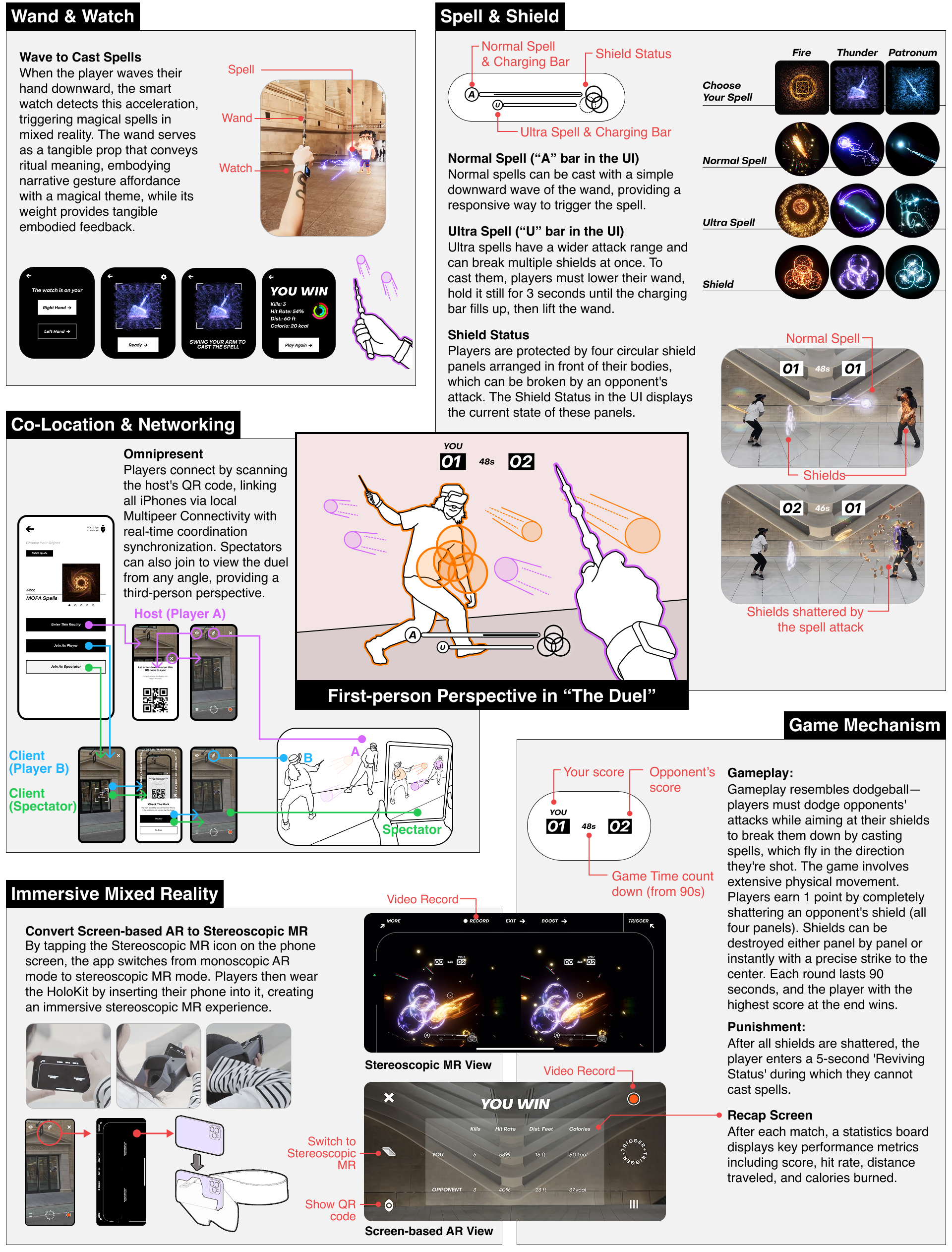}
    \caption{Detailed game and user experience design, using ``The Duel'' as an example. It illustrates: (1) the first-person perspective of the game with all UI elements users can view through the MR interface;
(2) interaction design details using Apple Watch  to detect spell-casting gestures;
(3) magic spell and shield system design;
(4) co-location and networking user flow: how two players connect and how spectators join to watch in third-person view;
(5) detailed round-based dueling game mechanism design; and
(6) conversion process from monocular screen-based AR view into immersive stereoscopic MR view based on HoloKit.}
    \label{fig:gamefeature}
\end{figure}

MOFA includes three co-located bodily game modes. Gameplay specifications are provided in Table~\ref{tab:gameprobes}. Detailed game and user experience design are depicted in Fig. \ref{fig:gamefeature}.

{
\renewcommand{\arraystretch}{1.5}  % Only applies inside this block
\begin{table*}[ht!]
\caption{Specifications of three MOFA game modes}
\resizebox{1\linewidth}{!}{%
\begin{tabular}{p{4.5cm} p{4.5cm} p{4.5cm} p{4.5cm}}
\toprule
\textbf{Gameplay} & \textbf{The Training} & \textbf{The Duel} & \textbf{The Dragon} \\
\midrule
\textbf{Required Number of Players} & 1 & 2 & 3 or more (with one player controlling the Dragon via a mobile device) \\
\textbf{Duration per Round} & 90 seconds & 90 seconds & Approximately 5 minutes \\
\textbf{Scoring Mechanism} & Cast magic spells to break the AI opponent's shield & Cast magic spells to break the opponent's shield & Cast magic spells to hit the dragon\\
\textbf{Winning Condition} & Achieve a higher score than the AI opponent & Achieve a higher score than the human opponent & Defeat the dragon by reducing its HP to zero \\
\textbf{Minimum Space Required (length x width x height)} & 1 × 2 × 2.5 m & 3 × 3 × 2.5 m & 5 × 5 × 3 m \\
\bottomrule
\end{tabular}%
}
\label{tab:gameprobes}
\end{table*}
}

\subsubsection{``The Training'' - a single-player gameplay mode}

``The Training'' is a single-player dueling mode where players engage in magical battles with an AI-controlled cartoon opponent (see Fig.~\ref{fig:gameplay}a). Inspired by Harry Potter's training scenes\footnote{\url{https://harrypotter.fandom.com/wiki/Cast-a-Spell_training_room}, visited on October 25, 2024}, players take on the role of wizards, honing their wand skills to cast spells and attack virtual ghosts in Hogwarts. This game probe focuses on one spell-casting interaction: detecting ``waving down'' hand gestures with physical wands. It is a single-player MRHMD gameplay in public spaces. 

\paragraph{Game flow of ``The Training''} First, the player puts on the Apple Watch while holding a wand and launches the game app on both the watch and iPhone. After selecting ``The Training'' mode, the player inserts the iPhone into the headset. Once wearing the headset and detecting a floor plane, the player taps the Apple Watch to place the virtual AI cartoon avatar opponent on the floor in front of them. Upon the audio cue ``1, 2, 3, start!'' the round begins, and the AI starts casting spells and dodging. The player dodges incoming spells and counterattacks by waving their watch-worn hand, which triggers spells through motion detection. Players earn one point by completely breaking their opponent's mixed reality shield. The player tries to break the AI opponent’s shield as many times as possible within 90 seconds; the player wins if their score exceeds the AI's.

\subsubsection{``The Duel'' - a competitive gameplay mode}

``The Duel'' is a competitive gameplay mode pitting two players against each other in magical combat. It expands on ``The Training'' mechanics, transitioning from player-versus-AI to player-versus-player (see Fig.~\ref{fig:gameplay}b). Drawing inspiration from Harry Potter's classic Wizard Dueling\footnote{\url{https://harrypotter.fandom.com/wiki/Duelling}, visited on October 25, 2024}, players use wands to cast spells and fight each other, incorporating dodgeball-like body movements to avoid attacks. This simulates future scenarios where MRHMD users might encounter each other on the street for impromptu duels.

\begin{figure*}[ht]
    \centering
        \includegraphics[width=1\linewidth]{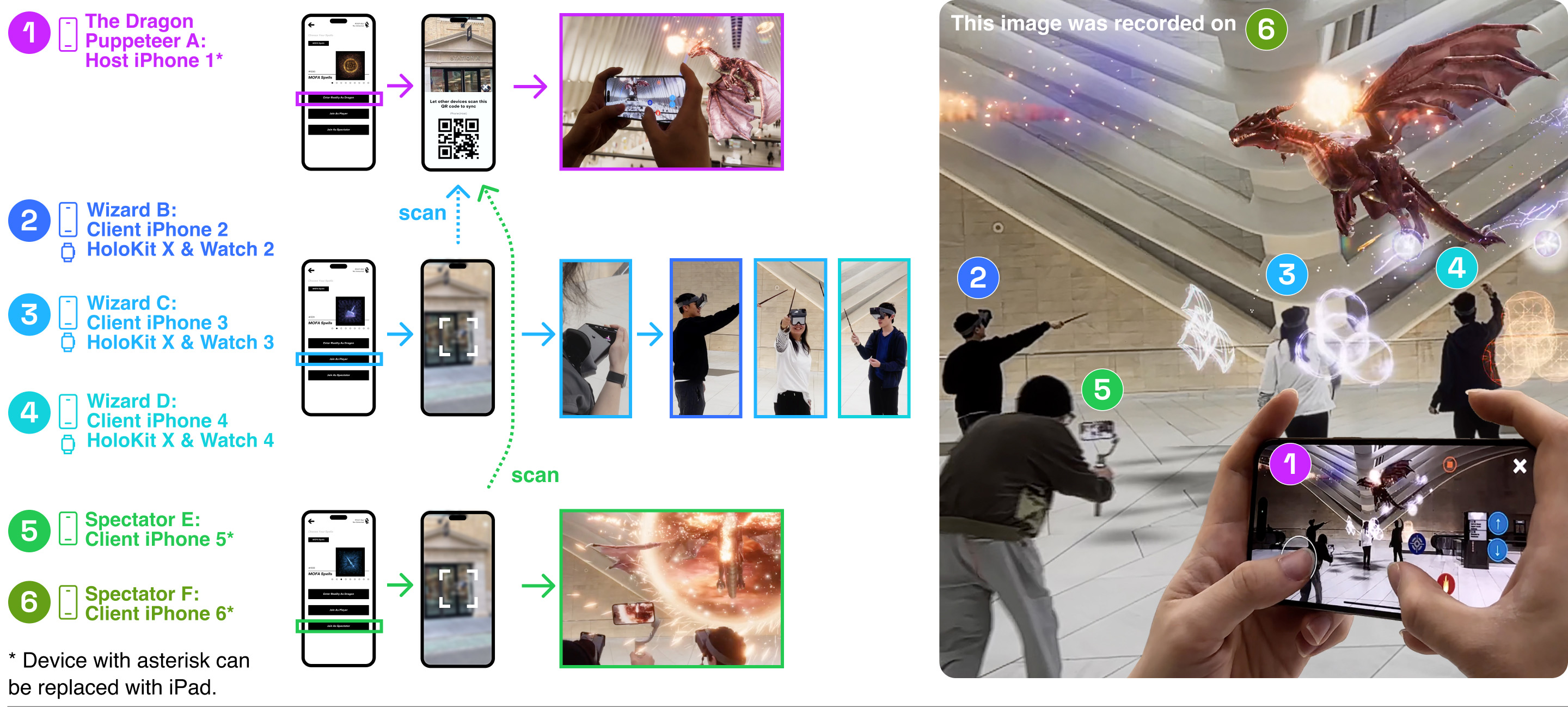}
        \par\vspace{1em}
    \includegraphics[width=1\linewidth]{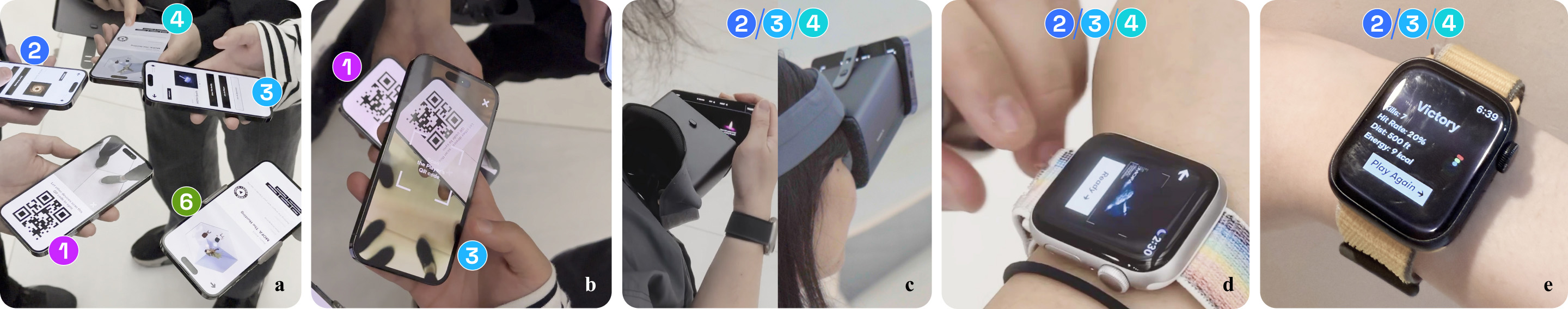}
    \caption{Top: Device-pairing diagram for MOFA ``The Dragon''. The image on the right shows a frame from a spectator video recorded on Device 6. Bottom: Step-by-step onboarding process: 
(a) All participants launch the game app on their iPhones.
(b) The host device initiates the game session, and guest devices join by scanning a QR code displayed on the host’s screen.
(c) Wizard players switch to stereoscopic mixed reality mode and insert their phones into the headsets.
(d) Wizard players press the ``Ready'' button on their watches to activate their powers and enter the game.
(e) After each round, performance data---``Score'' (kills), ``Hit Rate'', ``Distance'' (distance moved), and ``Calorie'' (calories burned)---are displayed on each wizard player's Apple Watch. Players can tap ``Play Again'' on the watch to start a new round. }
    \label{fig:userflowdragon}
\end{figure*}

\paragraph{Game flow of ``The Duel''} First, two human players launch the game app on both their watches and iPhones. The host player selects ``The Duel'' mode. A QR code is displayed on the phone. The client player scans the host's QR code with their iPhone. After confirming the connection, both players insert their iPhones into their headsets. Once wearing the headsets, the two players face each other and click their Apple Watches to start the game round. Upon the audio cue ``1, 2, 3, start!'' the players cast spells, dodge, and counterattack. Players earn one point by completely breaking their opponent's mixed reality shield. Each round lasts 90 seconds, and the player with the most shield breaks wins. 
After each round, players can check their Apple Watch to see their score, hit rate, estimated distance moved, and estimated calories burned. We observed that players often make dramatic movements to avoid being hit. Thus, we selected 90-second rounds to create a mild exertion game suitable for players with average fitness levels, burning approximately 10 kilocalories per round (see Fig.~\ref{fig:userflowdragon}e). %Note that Apple Watch calorie estimates are not accurate enough for scientific cardio research. 

\subsubsection{``The Dragon'' - a collaborative and puppeteering gameplay mode}
``The Dragon,'' designed for three or more players and extending from ``The Duel'',  introduces a new interaction paradigm (see Fig.~\ref{fig:gameplay}c): puppeteering. In this mode, one handheld AR player---dubbed the ``puppeteer''---controls a virtual dragon using their mobile phone, interacting with MRHMD users. Meanwhile, players equipped with MRHMDs become dragon-hunting warriors, casting spells to battle the dragon while using body movements to avoid attacks. This gameplay is inspired by Game of Thrones' famous dragon‑hunting scenes\footnote{\url{https://game-of-thrones-winter-is-coming-game.fandom.com/wiki/Dragon_Hunt}, visited on October 25, 2024}. It envisions a transitional future where MRHMD users coexist with handheld AR users, enabling collaborative play between different device types when they encounter each other on the street.

\paragraph{Game flow of ``The Dragon''} The host player puppeteers the dragon on their mobile screen (Device 1 in Fig.~\ref{fig:userflowdragon}). First, the host selects ``The Dragon'' mode and displays a QR code on their phone screen. Three human players (Devices 2, 3, 4 in Fig.~\ref{fig:userflowdragon}) scan the host's QR code with their iPhones (see Fig.~\ref{fig:userflowdragon}a-b). After confirming the connection, all players insert their iPhones into their headsets (see Fig.~\ref{fig:userflowdragon}c). Two spectators (Device 5 and 6 in Fig.~\ref{fig:userflowdragon}) may also scan the QR code with their iPhones to join as spectators. Once wearing the headsets, the three players face the designated play area after confirming with the host player and tap ``Ready'' on their Apple Watches to indicate readiness (see Fig.~\ref{fig:userflowdragon}d). The host uses their phone to detect the floor plane, then spawns the virtual dragon to begin the battle. Upon the audio cue ``1, 2, 3, start!'', the round begins. Players cast spells to attack the dragon while dodging its counterattacks. The host controls the dragon via touchscreen input on their phone, using navigation similar to that of drone driving. They can also launch fireballs at players by locking onto them on the mobile screen. Similar to the gameplay of ``The Duel'', players often make dramatic movements to avoid being hit. The goal is to defeat the dragon by reducing its health points (HP) to zero. Each round lasts approximately 5 minutes---if time expires, the dragon wins; if the dragon's HP reaches zero, the wizards win.

\subsection{System Implementation}

To address the cost barrier of MRHMDs noted in Section \ref{reason:inaccessibility}, we employed HoloKit\footnote{\url{https://holokit.io/products/holokit-x}, visited on October 25, 2024}, an affordable headset (US\$129) \cite{Hu2024HoloKit}, which is a smartphone-based, open-source device providing a stereoscopic optical see-through mixed reality experience. This cost-effective solution facilitates multiplayer mixed reality experiences ``in the wild'', enabling numerous passersby to engage with the gameplay.

The game probes were implemented using Unity 6 LTS\footnote{\url{https://unity.com/releases/unity-6}, visited on October 25, 2024} and the HoloKit SDK\footnote{\url{https://github.com/holokit/holokit-unity-sdk}, visited on October 25, 2024} based on ARKit\footnote{\url{https://developer.apple.com/augmented-reality/arkit/}, visited on October 25, 2024}. We deployed the game as an iOS application on iPhone 12 or newer models, which are inserted into the HoloKit. The headset displays stereoscopic renderings on the iPhone's screen, allowing users to see stereoscopic 3D images overlaid on reality.

Interactions with augmented content were achieved through an Apple Watch. The watch's motion sensors capture wrist movements for spell-casting, enabling ranged attacking gameplay without physical contact between players. This approach sidesteps the physical contact issue by allowing players to cast magic spells for intangible interaction with opponents. MOFA game probes feature various spell classes, each comprising a basic attack spell and an ultra spell. Players select one magic class per game round. While basic attack spells have minor variations in speed and size, ultra spells differ significantly, adding depth to the gameplay.

\begin{figure}[ht]
    \centering
    \includegraphics[width=0.3\linewidth]{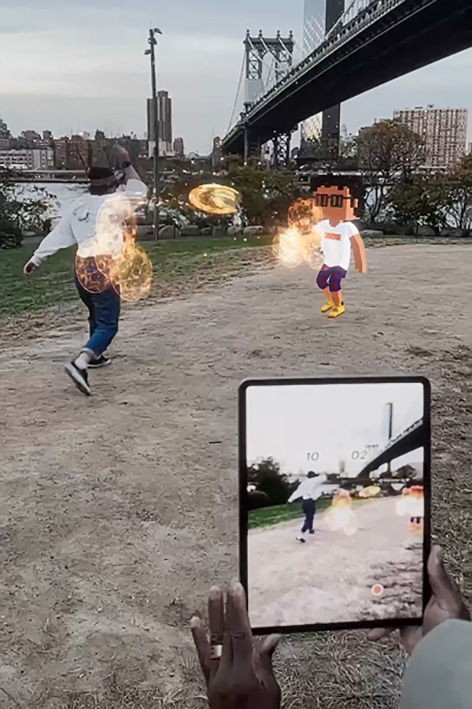}
    \caption{Spectators can use iPads to view the augmented reality experience from a third-person perspective and capture recordings of what they see.}
    \label{fig:spectator}
\end{figure}

We connected all co-located devices using the InstantCopresence toolkit~\cite{hu2023InstantCopresence}. The toolkit enables spontaneous mixed reality sessions when users scan a QR code on the host device. QR code scanning enables device co-location by calculating the relative position of client devices to the host's QR code, allowing multiple devices to sync their coordinate systems based on their positions relative to the host origin. The network layer uses Multipeer Connectivity---Apple's peer-to-peer local networking technology that powers AirDrop---to provide local networking with latency under 20 ms. This enables street play anywhere, even in rural areas without cellular or Wi-Fi coverage. An auto-sync system keeps virtual object data consistent across devices, ensuring all participants see the same virtual elements. Spectators can view the augmented content from a third-person perspective using mobile devices like iPads (see Fig.~\ref{fig:spectator}).

% utilizing a smartphone-based open-source mixed reality headset, HoloKit\footnote{https://holokit.io/products/holokit-x}, which is inexpensive for multiplayer mixed reality. 

  % amber -->
\section{Method}
\label{sec:method}

% \usepackage{graphicx}

% 在这里写大的method
% research through design
% 混合method： desktop research and in the wild

% Using an in-the-wild research method, we observed the reactions of both players and passersby in public spaces. To gain a broader perspective on user reactions, we also released video recordings of the street play online. We then conducted desktop research, gathering and analyzing online reviews and comments to assess the sentiments expressed.

To investigate IMRSP, we conducted an in-the-wild study~\cite{Benford2013Performance} combining direct gameplay video observations with immediate post-play contextual inquiries \cite{Duda2020Contextual}. To gain a broader perspective, we shared in-the-wild gameplay recordings on a tech‑enthusiast forum. Following this, we conducted desktop research to gather and analyze online reviews and comments about the experience. This combined approach allowed us to capture both firsthand player experiences and wider public perceptions. Our study was approved by the authors’ institutional ethics review, and all participants gave informed consent prior to participation.

\subsection{In-the-Wild Field Study}

We conducted a series of in-the-wild play sessions over three years, in which members of the public were invited to try MOFA. These sessions occurred as pop-up events at 20 different locations across North America, Asia, Europe, and Oceania, spanning settings such as university campuses, tech expos, urban parks, public building courtyards, and natural wilderness sites. 

\subsubsection{Scenarios}

\begin{figure*}
    \centering
    \includegraphics[width=1\linewidth]{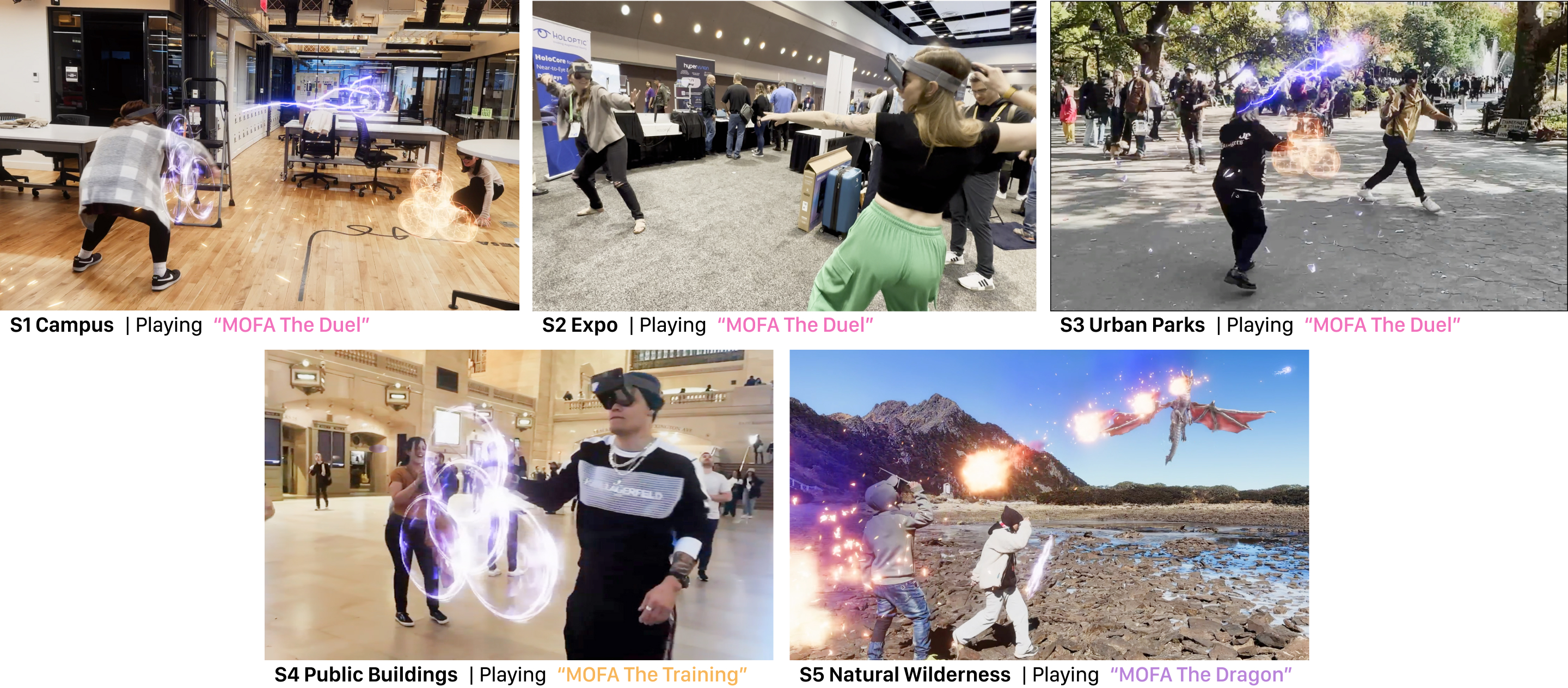}
    \caption{Five representative scenarios of volunteered participation and interviews. S1: Campus; S2: Expos; S3: Urban Parks; S4: Public Buildings; S5: Natural Wilderness. }
    \label{fig:scenario}
\end{figure*}

For this paper, we focus on five representative scenarios (see Fig.~\ref{fig:scenario}): public areas of university campus (S1), exhibition spaces at expos (S2), urban parks (S3), public building atria (S4), and scenic mountainside in natural wilderness (S5). These diverse contexts allowed us to examine IMRSP across varying degrees of publicness, crowd density, and thematic fit.

\textsf{S1 Campus}: Semi-public university courtyards have moderate foot traffic (5-10 passersby per session). The steady trickle of students and faculty creates a sociable backdrop for exploration. Studying IMRSP here reveals how ambient social presence affects curiosity and engagement with the game in semi-public familiar settings.

\textsf{S2 Expos}: Expo and conference halls present a high-visibility, demonstration‑friendly environment characterized by dense crowds (30-50 observers per session), rapid turnover of visitors, and visually busy booth layouts. IMRSP deployment here uncovers strategies for maintaining engagement amid rapid attendee turnover, where audiences are generally more curious and have a higher threshold for social acceptance. 

\textsf{S3 Urban Parks}: Public outdoor areas offer open space, greenery, and diverse pedestrian flow (10-20 bystanders per session), including Dumbo Park in Brooklyn and Washington Square Park in Manhattan. Testing in these environments shows how natural elements and unstructured layouts affect spontaneous participation in leisure settings.

\textsf{S4 Public Buildings}: This scenario is characterized by iconic indoor spaces with continuous pedestrian flows, echoic soundscapes, and high visibility (20-50 people per session), such as Oculus Transportation Hub and Grand Central Terminal. Gameplay in these settings highlights necessary design adaptations for embodied interactions under public social pressure. 

\textsf{S5 Natural Wilderness}: This is a typical scenario that has unbounded landscapes with open ground, vertical space, and minimal bystander presence (0-2 bystanders per session), such as the foot of mountains in national parks. Exploring IMRSP here reveals how the experience transforms when social constraints disappear and nature becomes the primary contextual element.

\begin{figure*}
    \centering
    \includegraphics[width=1\linewidth]{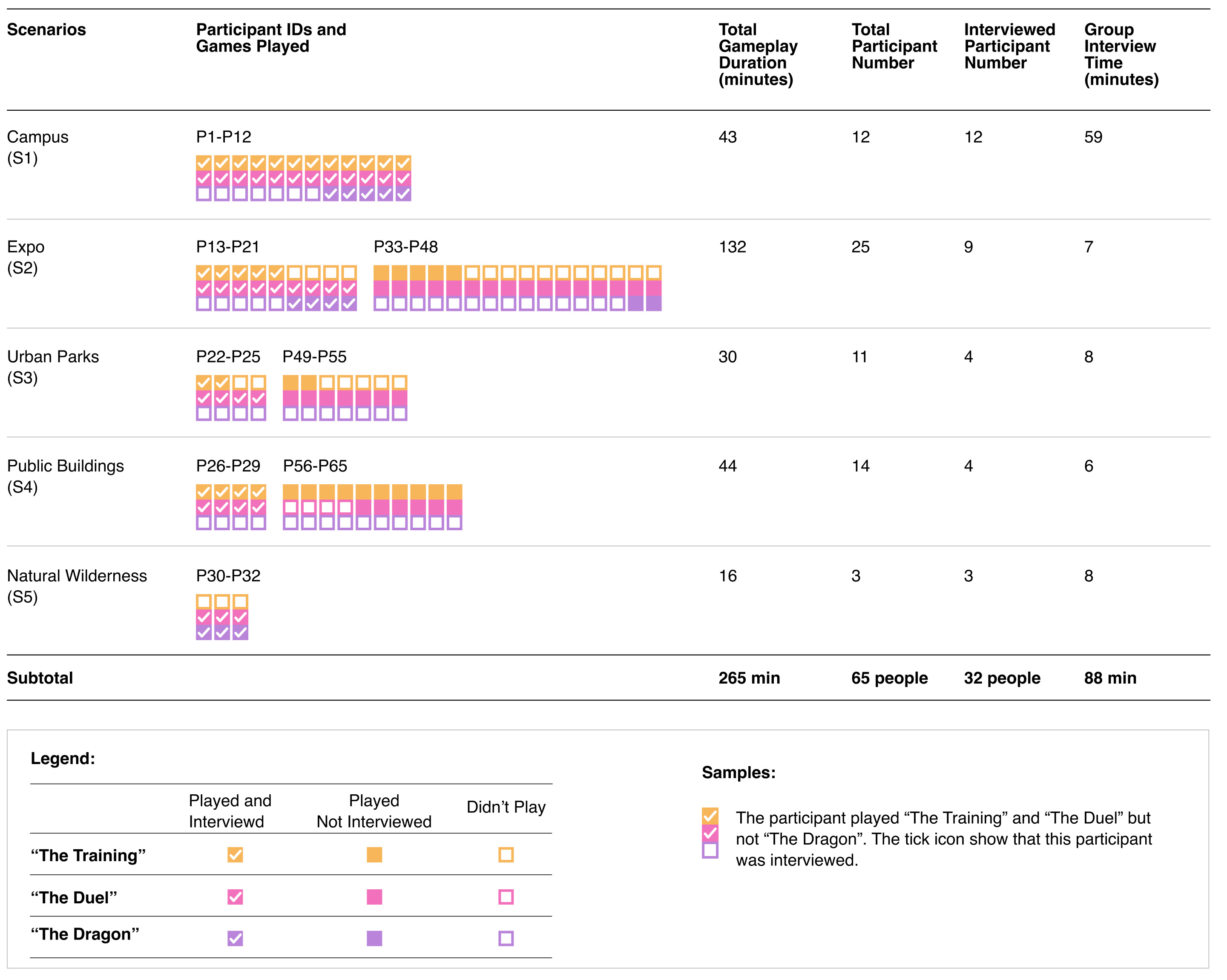}
    \caption{Data collection from the in-the-wild study.}
    \label{fig:datawild}
\end{figure*}

\subsubsection{Procedure}

% \leavevmode\vspace{0.1em}

In order to collect ecologically valid MR street play data, we used an opportunistic recruitment strategy, following the method used in previous works \cite{Denning2014situ,OHagan2021Safety}. This recruitment approach is common in in-the-wild research \cite{Rogers2017Research}, which yielded relatively short engagements due to the nature of interrupting people during their normal routine. Nevertheless, it encompassed a broad spectrum of individuals who encountered this MR street play in those public spaces, thereby enhancing the diversity of our sample. The procedure for a typical street‑play session is described below.

\begin{enumerate}

\item \emph{Configure the play area}: In each pop-up or in-booth show, two or three researchers ran the session in a space of at least 3 × 3 meters in a public area, ensuring minimal disruption to pedestrian flow. A ``Video Recording In Progress'' sign was displayed. For Expo scenarios (S2), we demonstrated the game at the booth and via pop‑ups in lobbies or near entrances during breaks, following our in‑the‑wild deployment approach.

\item \emph{Engage passersby}: Pedestrians were drawn to the ongoing gameplay, introduction videos, and spectator footage displayed on iPads and portable screens. As passersby were drawn to the play and began to interact with the researchers, we invited them to try the game.

\item \emph{Obtain informed consent}: Interested participants signed consent forms on an iPad or paper.

\item \emph{Equip participants for a game mode}: To facilitate a smooth and context-aware play experience, researchers assisted participants with initial setup, including pairing players' devices, wearing the MR headset, securing the motion-tracking watch, and selecting their preferred virtual spells. The choice of MOFA game mode was guided by a combination of player preferences, social context, site constraints, and time availability (see Table~\ref{tab:gameprobes}). Participants were introduced to the available game options, and the most suitable mode was recommended based on the situation. The adaptive approach maintained ecological validity while ensuring that each participant could meaningfully engage within the constraints of the public and often time-limited settings.

\begin{itemize}
    \item Single participants typically began with ``The Training'', a one-player mode designed for onboarding and individual practice. 
    \item When another participant became available, they were often invited to join in ``The Duel'', which requires two players to engage in competitive interactions. Participants who arrived together---such as friends or colleagues---were directly introduced to ``The Duel'' to leverage their existing social rapport. 
    \item When space permitted extended movement (e.g., campus playgrounds or natural wilderness) and participants had the time and physical ability, researchers suggested ``The Dragon'', a longer (about 5 minutes) and more physically demanding cooperative game mode. 
\end{itemize}  

\item \emph{Facilitate gameplay sessions}: The first 90 seconds (the first round) served as a tutorial. Once participants learned to attack and dodge in MR, they played one or more additional rounds. 
After each round, participants could view results in a summary chart in MR view or on Apple Watch, including score, hit rate, distance moved, and calories burned.
    
\item \emph{Conclude gameplay}: Participants could play for their desired duration. However, as the MOFA games involved vigorous physical activity, players typically concluded after 3--5 minutes due to physical fatigue from bodily exertion. When bystanders wished to join, researchers rotated them into the session. This fluid grouping resembled real-world street play, where people can spontaneously join or leave a game. 

\item \emph{Contextual inquiry}: Researchers immediately asked participants if they had time for a brief interview. If they declined, the researchers let them leave. Otherwise, one researcher interviewed either individuals or player groups while another moderated gameplay. During the interview, participants shared their experiences and thoughts either one-on-one or in their original game groups of 2--3 people. Although brief, these interviews captured valuable immediate reactions while participants could still vividly recall their fresh experience. We used a prepared set of open‑ended questions listed in Appendix \ref{appendix:interview}. 

\end{enumerate}

\subsubsection{Data Collection}

For the in-the-wild observations and post-play interviews, two types of video recordings were collected: AR spectator views and interview recordings.
For each session, a researcher held an iPad with a spectator view, which shows the third-person perspective of player behaviors with augmented views of game content (see Fig.~\ref{fig:spectator}). In total, we collected 265 minutes of recordings of players' behavior data. In addition, we collected a total number of 88 minutes of post-play interview video, as detailed in Fig.~\ref{fig:datawild}. 

% \added{For gameplay video, a researcher used an iPad to run a special spectator mode that shows a third-person view of the scene with virtual game elements composited. Essentially, the iPad camera records the players moving and reacting, while also showing the spells and effects as if one were seeing the AR through a camera lens. This provided rich behavioral footage for later analysis without needing a complex external camera rig. In total, we recorded 265 minutes of gameplay video across all sessions covering 65 distinct players (\ref{fig:datawild}). 

% The total lengths of the final recordings used for the research include only actual gameplay. To optimize coding and file size, we excluded activities like greetings, introductions, consent signing, technical setup, tutorials, and breaks between gameplay sessions. Thus, these times reflect pure gameplay, player reactions, or interviews. 

% \paragraph{Interview recordings with players and spectators.} We collected a total number of 88 minutes of post-play interview video (see Fig.~\ref{fig:datawild}). 

% Like the gameplay video, the total lengths of the final recordings used for the research include only actual interview conversations and didn't include greetings or intervals.

The contextual nature of these interviews allowed us to capture situated, embodied reflections in the moment, while the prepared questions list ensured consistency across sessions. Due to the fast-paced and public setting, especially in high-traffic environments like Expo (S2), Urban Parks (S3), and Public Buildings (S4), most participants only had a limited window for interaction, resulting in short interview durations (typically 2--4 minutes). Some participants who had consented to participate ultimately declined the interview due to time sensitivity or discomfort with extended conversation in public. We respected these preferences and included only the data of those who voluntarily continued. This pragmatic approach ensured ethical engagement while still allowing for the collection of rich, immediate feedback from diverse real-world contexts.

% The interviewed participants were also those prioritized participants who achieved high scores, demonstrated extended engagement, or exhibited diverse behavioral patterns and demographic backgrounds. In total we gathered 88 minutes of interview recordings from 32 on-site participants (Table 1).

% 
% During gameplay sessions, Researcher A provided instructions to participants, while Researcher B documented participant reactions using smartphones equipped with mixed-reality spectator views. Following each session, Researcher A approached participants for potential interviews. With participant consent, Researcher B conducted semi-structured interviews lasting 5-15 minutes, either individually or in small groups of 2-3 players who had participated together. All interviews were video-recorded. The selection of interviewees prioritized participants who achieved high scores, demonstrated extended engagement, or exhibited diverse behavioral patterns and demographic backgrounds.

% \subsubsection{Participants}

Over the study period, 65 individuals played the game across all sessions, of whom 32 volunteered for post-play interviews. These participants (labeled P1--P32 for those interviewed, and P33--P65 for additional players, who did not participate in the post-play interviews) represented a broad demographic spectrum. The video data indicated a roughly balanced gender mix and a wide range of ages (teens with parents present, college students, adults, and a few seniors) and ethnic backgrounds. 

% The context of each scenario influenced the participant pool: for example, in the campus deployment (S1), many players were undergraduate students; at the tech expo (S2), most were tech industry professionals or enthusiasts; in parks and public spaces (S3, S4), players came from varied occupations and walks of life who happened to be passing by; in the mountain/wilderness scenario (S5), participants were hikers and tourists. We did not collect detailed personal profiles for anonymity and ethical simplicity (given the ad-hoc recruitment), but the diversity of locales inherently gave us a heterogeneous sample of users.
% All participants who were interviewed are referenced with codes P1--P32 in our results. 
% When quoting or describing behaviors of those who did not do an interview (but were observed), we refer to them in aggregate (e.g. noting X out of 65 players did a certain action). 

\subsubsection{Safety and Ethical Considerations}
We avoided any hazardous locations. Although privacy is not expected in public spaces, we followed strict ethical guidelines during data collection. Our recordings focused on gameplay, and we avoided recording people unnecessarily. Participants or bystanders could opt out before interviews. 
No minors were interviewed; a few minors who tried the game did so under parental supervision, but we did not include their feedback in our dataset beyond general observations.

\subsection{Desktop Research of Online Forum Reviews}

\begin{figure*}
    \centering
    % First subfigure
        \includegraphics[width=\linewidth]{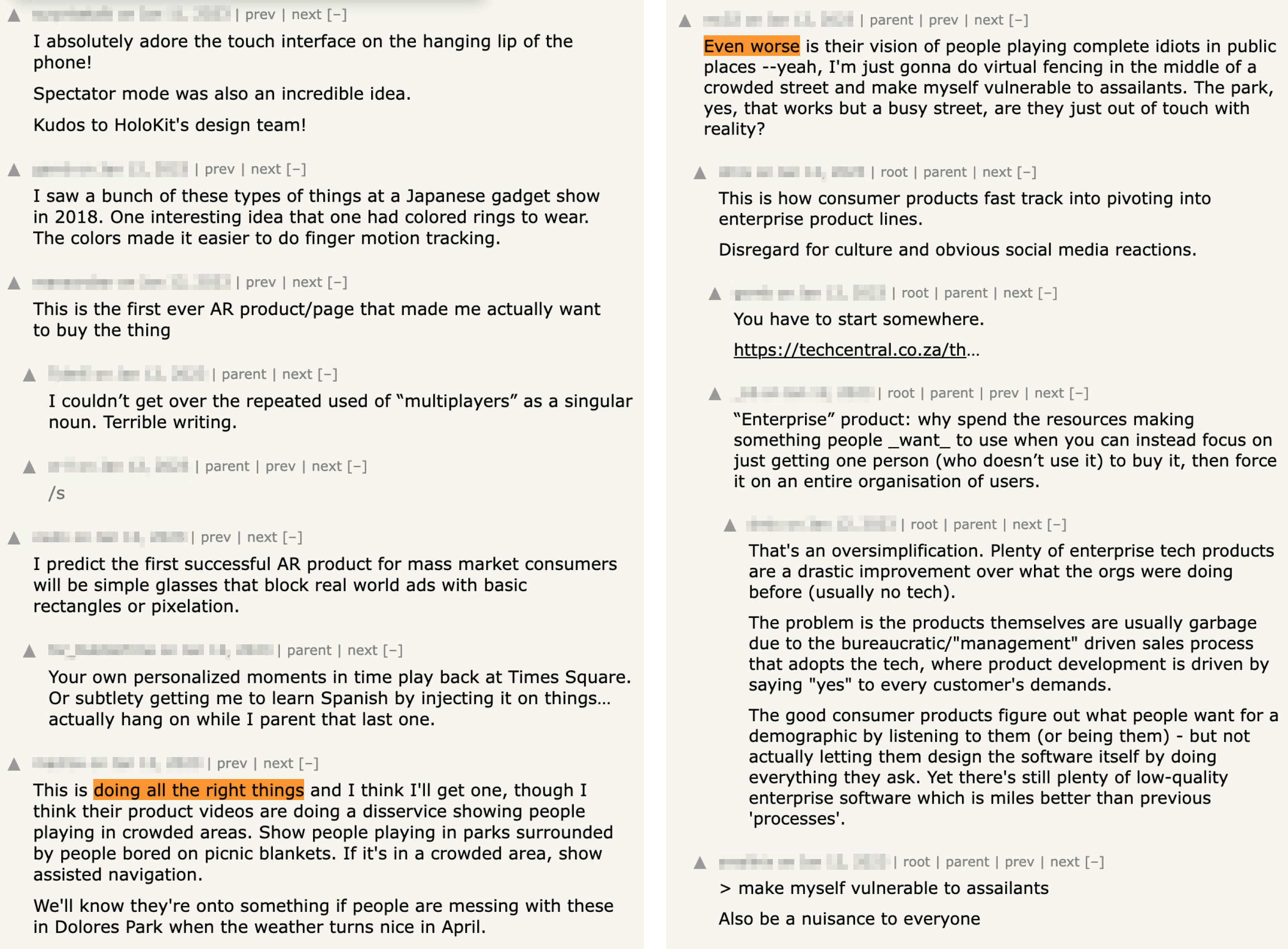}
    \caption{Comments on actual MOFA game video recording posted on an online tech-enthusiast forum. }
    \label{fig:onlineforum}
\end{figure*}

In addition to on-site data, we wanted to capture broader public perceptions of IMRSP---including those of people who had not played the game but saw it online. To do this, we conducted a desktop analysis of online discussions about our game probe. We posted a short gameplay video of MOFA and a description of the concept on a popular tech-enthusiast forum, inviting feedback and discussion. Over several weeks, this thread garnered considerable attention and commentary. We collected all responses from this forum thread, resulting in 14,919 words of raw text across numerous comments. Fig.~\ref{fig:onlineforum} shows snippets from the online forum threads. We then filtered this dataset to focus on relevant content, specifically, comments that addressed the social implications, user experience, or concerns about MR street play (as opposed to off-topic remarks about technology or economics). After filtering out unrelated discussion, we had 1,325 words of substantive comments from 31 unique online respondents (L1--L31). These comments ranged from short quips to multi-sentence opinions (5 to 204 words each).

The online respondents had all viewed the same gameplay footage that our real-world participants had experienced (via the posted video). Their perspectives are thus those of observers envisioning the experience, rather than firsthand players. Not being physically present or involved with researchers, their feedback was not influenced by any demand characteristics or desire to please us. This provided us with more candid feedback and skepticism than we typically observe on‑site. Some commenters were enthusiastic about the idea, while others were critical or concerned. We treated this as an additional qualitative dataset that complements the interview data: on-site participants provide insight into actual experience and immediate personal feelings, whereas online responses reveal more reflective or hypothetical viewpoints, including critiques from a social acceptance standpoint.

% \begin{table*}
%     \caption{Data collection from the online forum }
%     \centering
%     \includegraphics[width=1\linewidth]{images/table_online.jpg}
%     \label{tab:datadesktop}
% \end{table*}

\subsection{Data Analysis}

All interview video recordings were transcribed, and field notes from gameplay video observations were expanded into textual descriptions. For the online forum data, we already had the comments in text form. We analyzed the combined qualitative data using a content analysis with emergent thematic coding. We did not start with a rigid codebook; instead, three researchers independently read through subsets of the data (including transcripts of the observation and interview videos and online comments) to inductively identify recurring themes and patterns. After an initial open-coding pass, we compared notes and converged on a set of candidate themes capturing salient aspects of the experience. We then iteratively refined these, grouping related subthemes and ensuring they were grounded in the data.

During coding, we were careful to differentiate between on-site participant responses (P-codes) and online reviewer comments (L-codes). We created separate code categories when needed to reflect differences in perspective. 

% For example, ``awkwardness'' was noted by both groups, but online commenters raised it more frequently and harshly; thus, we coded occurrences in the L data and P data and later compared them. We also quantified some occurrences to get a sense of prevalence (e.g., how many of the 63 total contributors mentioned a certain issue). These counts (noted as, say, 13/63) are used in the results to convey magnitude but are not treated as statistical measures. 

Throughout the analysis, we practiced researcher triangulation: multiple coders cross-checked each other’s assigned codes and interpretations. Differences were discussed and resolved. We also revisited raw videos to verify behaviors for certain themes. This process ensured that our findings are solidly grounded in the empirical evidence. We group our findings by four broad thematic categories: (1) Social Interactions, (2) Social Concerns, (3) Gameplay Experience, and (4) Environment and Location. 

% One researcher transcribed all interview recordings. Three researchers initially coded one video and one interview from each venue and online materials collectively to establish baseline agreement. One researcher then coded all materials comprehensively, generating 121 unique codes organized into five major themes, as detailed in the findings section. Three researchers then collaboratively reviewed and adjusted the themes, reaching a strong inter-coder reliability.

 % elan -->
\section{Findings}
\label{sec:findings}

% \usetikzlibrary{arrows.meta, matrix}
% \begin{tikzpicture}
% \matrix (lang) [matrix of nodes, column sep=3mm, row sep=3mm,
%     nodes={anchor=center, minimum width=2cm, minimum height=2cm, rounded corners=3mm,
%         fill=cyan!90!black, draw=cyan!50!black, line width=.5mm}]
% {F\# & Swift\ref{intent:m5} &  A \\
% Fortran & Java & A \\};

% \draw [->, shorten >=-3mm](lang.south west)--(lang.south east); 
% \node[below] at (lang.south-|lang-1-1.center) {MR HMD}; 
% \node[below] at (lang.south-|lang-1-2.center) {Handheld AR};
% \node[below] at (lang.south-|lang-1-3.center) {No AR Device};

% \draw [->, shorten >=-3mm](lang.south west)--(lang.north west);

% \node[rotate=90, above] at (lang.west|-lang-2-1.center) {low};
% \node[rotate=90, above] at (lang.west|-lang-1-1.center) {high};

% \draw[dashed] (lang.south) node[below=5mm, font=\bfseries\sffamily]{Device Diversity}--(lang.north);
% \draw[dashed] (lang.west) node[rotate=90, above=5mm, font=\bfseries\sffamily]{Intention to involve} --(lang.east);
% %\node

% \end{tikzpicture}

In this section, we present our findings, which are organized into four key themes that emerged from the data analysis. These themes capture different facets of the IMRSP experience and provide a framework for understanding our findings. We complement our qualitative insights with behavioral coding, using the notation $x/y$ to show that $x$ out of $y$ participants demonstrated a specific behavior or attitude. Specifically, $y=65$ for on-site participants' behaviors (P1--P65); $y=63$ for participant and respondent attitudes (P1--P32, L1--L31). We link quotes with personal IDs to indicate whether the insights come from on-site participants (e.g., P1) or online respondents (e.g., L1).

\subsection{Social Interactions}

We observed that players quickly formed connections with each other during play, and the presence of an audience introduced performative elements. The game blurred the line between players and spectators, creating new opportunities for engagement in a public setting.

\subsubsection{Collaboration and competition (31/65)}
\label{theme:collaboration}

IMRSP facilitates the establishment of social connections and fosters a positive community atmosphere through both competition and collaboration. Because the game is played face-to-face in a public space, it enables direct, real-time interactions among players. We observed many collaborative and competitive behaviors that strengthened bonds between players. For example, participants often taught and explained the gameplay to newcomers, encouraged their teammates with exclamations like \dquote{come on!} or \dquote{one more shot!}, warned each other of the dragon’s movements by shouting \dquote{watch out!}, and celebrated victories together. Celebratory acts included shouting \dquote{Yes, we did it!} after defeating the dragon, congratulating each other with double high-fives or fist-bumps, and even engaging in friendly taunts like \dquote{I will beat you!} during matches.

In interviews, participants consistently favored the multiplayer aspect of the game, noting that competition and collaboration both strengthened social ties. They emphasized the unique excitement of playing together in person. One player explained, \dquote{If there are many people, and there is a certain sense of excitement and competition} (P4), suggesting that having more people involved made the experience more thrilling. Others simply enjoyed this way to spend time with friends and family, remarking that \dquote{Anyway, it’s interesting to have a new way to play with people} (P27). As another participant put it, \dquote{Probably the interesting part is not the magic, but the person on the opposite side} (P10), highlighting that the real fun came from interacting with other people rather than the technology itself.

\subsubsection{Showmanship and performance (25/65)}
\label{theme:showmanship}

Because IMRSP unfolds in public and is highly visible, it often encourages participants to show off and perform. The game provides a unique opportunity for adults to engage in playful, out-of-the-ordinary behaviors in a public setting, which draws attention from bystanders and creates a shared sense of excitement. In the presence of an audience, players sometimes embellished their actions with dramatic flourishes purely for entertainment, even if these gestures did not improve their chances of winning. We observed participants incorporating theatrical body movements into their play, thereby contributing to the overall spectacle for themselves and onlookers. Some examples of this playful showmanship include:

\begin{itemize}
\item Jumping (3/65)
\item Moving in a squatting stance (3/65)
\item Turning their back to the opponent (4/65)
\item Spinning around (2/65)
\item Clicking wands together with teammates or opponents as a ritual (13/65)
\end{itemize}

\begin{figure}
    \centering
    \includegraphics[width=1\linewidth]{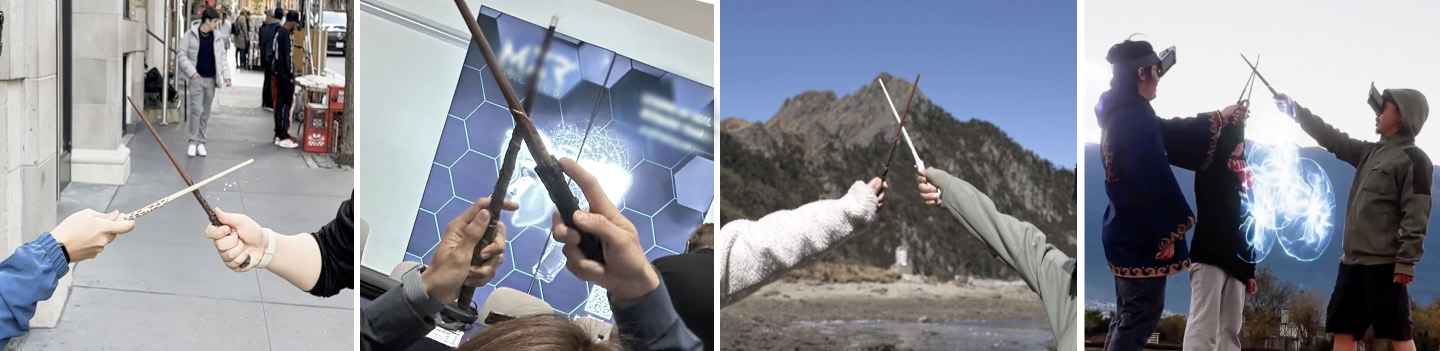}
    \caption{Players develop a ritual of playful showmanship: they click wands together with teammates or opponents before or after play, establishing a popular game tradition that resembles the wand interactions of wizards in Harry Potter movies.}
    \label{fig:clickingthewand}
\end{figure}

One particularly popular instance of showmanship was the ritual of clicking wands together before a match (see Fig.~\ref{fig:clickingthewand}). This gesture---featured in the game's tutorial video---quickly became a customary way for players to bond; many would click their wands together both before and after the game for photos or simply as a friendly greeting (P1, P2, P11, P12, P15, P16, P30--P32, P50--P53).

In some cases, just putting on the MR headset and holding the wands immediately put participants into a playful, performative mindset. Even before researchers officially started an MR session, players would begin to role-play and make exaggerated moves (P13, P14, P15, P59). Simply donning the gear was enough to inspire some to wave their wands dramatically or strike \dquote{magic-casting} poses, entertaining both themselves and anyone watching.

% The observable nature of the MR street play encourages showmanship among participants. The high-tech game provides adults with a unique opportunity to engage in non-daily behaviors in a public space, drawing attention from the audience and fostering a sense of excitement.

% In many observable activities, the presence of the audience and the overall excitement of the experience can inspire participants to incorporate dramatic body movements and posts into their performance. This may not necessarily enhance their chances of winning but rather serve to entertain themselves and the audience and add to the overall spectacle. Some examples of these showmanship elements include: 

% One particularly notable example is clicking wands together before the match. As this moment was captured in the tutorial video, this action quickly became a ritual for the game, with many players repeating the wand-clicking gesture before and after the game for photos or as a display of friendliness (P1-2, P11-12, P15-16, P30-32, P50-53) (see Fig. \ref{fig:clicking the wand}).

% In some instances, the mere act of wearing the headset and wielding the wands was enough to inspire participants to engage in role-play and dramatic body movements, even before the researchers had officially started the MR session (P13, P14, P15, P59). 

\subsubsection{Spectating and honeypot effect (41/65)}
\label{theme:spectating}

\begin{figure}
    \centering
    \includegraphics[width=0.8\linewidth]{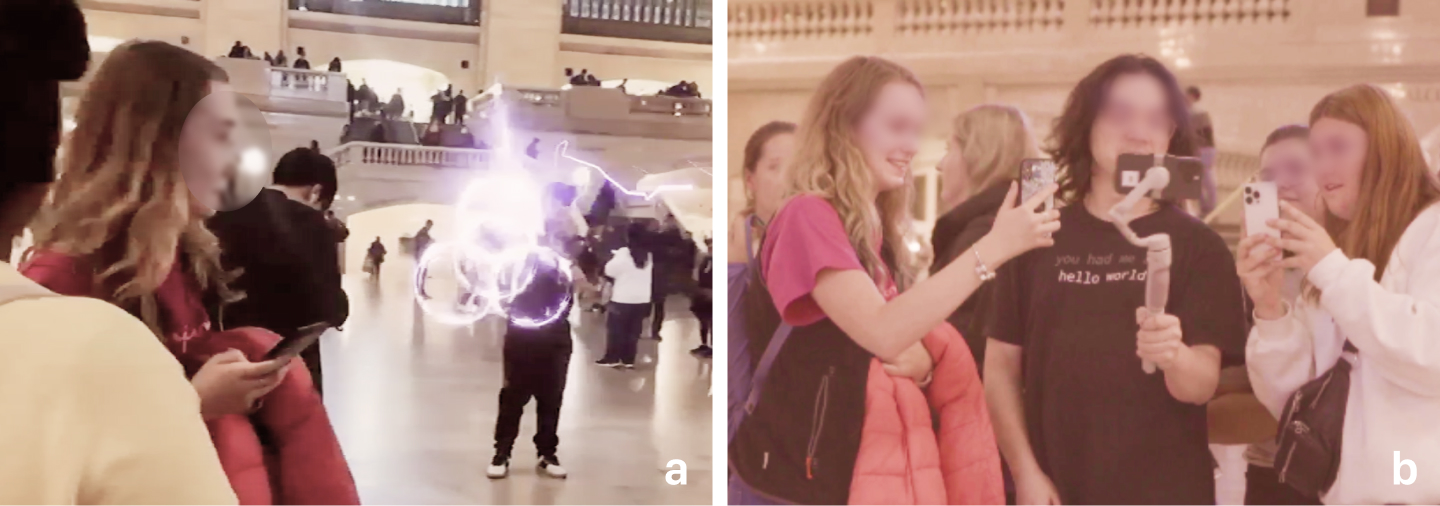}
    \caption{Honeypot effect observed: (a) Passersby were attracted by players' body movements, then (b) approached the researcher holding the spectator view of the game, watched the augmented game view, and took pictures of it. }
    \label{fig:passersby}
\end{figure}

When played in public spaces, MR street play invokes strong interest from passersby as it is fully observable, and arguably hard not to notice.
People stop to spectate for various reasons, including being attracted by the player's body movements, the spectators' view, or being attracted by other observing people (see Fig.~\ref{fig:passersby}). Typical observed behaviors of the spectators include: 
\begin{itemize}
\item Taking photos or videos of the players during the game (12/65)
\item Recording the spectator view screen to capture the mixed-reality action (7/65)
\item Cheering for the players (25/65)
\item Sharing photos or videos of the scene with friends or on social media (7/65)
\item Chatting with other bystanders about the game (34/65)
\item Volunteering to join the next game session themselves (41/65)
\end{itemize}

Participants expressed increased curiosity and desire to join the game after seeing the spectator view, with comments like \dquote{After I see the spectator view..., I really want to join} (P10) and \dquote{This is sick! I want to play too} (P31). In addition, the online reviews highlight the importance of the spectator mode, with one comment noting, \dquote{spectator mode was also an incredible idea} (L14). 

The spectator view was also found to add a social and enjoyable element to the game, as one participant (P3) mentioned, \dquote{If they have the spectator view, the bystanders can also comment on the game when the two of us are dueling, this is more fun.} However, not having access to the MR view can be detrimental to the experience, as P9 noted, \dquote{If he can't see MR at all, he's actually an obstacle in my game for me.}

\subsubsection{Inclusive multiplayer with handheld mobile device (8/65)}
\label{theme:joining}
The controller role on handheld devices turns out to be more popular than the MRHMD player roles, despite many participants not having tried the game in HMD before. 
When the researchers showcased the dragon hunting game, all players and spectators eagerly volunteered to be the dragon controller, even if they had never used the HMD or played the wizard duel game yet. One participant stated \dquote{Being the boss behind the scenes sounds totally sick} and \dquote{I wanna blast my friends with some fire!} (P32). Another shared their surprise at the addition of the controller role, stating, \dquote{Fight the Dragon gave me a new AR experience, because I used to think only of two-player battles, and I didn't expect that there would be a third machine to control the character} (P18).

\subsubsection{Appreciation of AR over VR (8/63)}
\label{theme:appreciation}

People expressed a strong preference for augmented reality experiences over virtual reality, both on-site and online. The ability to interact with friends and other facilities in the real world, rather than fully immersing in a virtual environment, was a key factor in their enjoyment and interest in the game. As one online comment stated, \dquote{I am way more interested in a digital world being overlaid on the real world---augmenting rather than escaping} (L22). Other participants highlighted the limitations of VR, mentioning, \dquote{You can't even pick up a beer while playing VR} (L18); \dquote{You can't see around you in Quest unless you tap on the side... personally, I like to be immersed in the game while also being able to see what's around} (P19). 

\subsection{Social Concerns}

While the social benefits of MR street play were evident, our study also uncovered significant social concerns and friction points, particularly regarding how this novel activity is perceived within the broader society. These concerns were more frequently and vividly expressed in the online reviews, though some on-site participants acknowledged them when prompted. We detail these issues below.

\subsubsection{Social awkwardness and perceived impropriety (13/63)}
\label{theme:awkwardness}

Thirteen respondents (13/63), all from the online forum data collection, expressed feelings of awkwardness as their main impression of the game, with some feeling offended by the prospect of engaging in immersive MR street play. These comments questioned the social acceptance level of wearing the device in public, particularly regarding the potential impact on non-participants. For example, \dquote{I'm not sure if I'd like to walk around the streets wearing them though---it would likely be incredibly annoying to other people} (L1). L19 criticized the play as \dquote{disregard for culture and obvious social media reactions.}
One comment expressed empathy for passersby who might be unintentionally captured in the middle of others' game space, noting that they \dquote{don't look too pleased} (L26). The impropriety of engaging in the game in certain contexts was also highlighted, with L10 commenting, \dquote{It seems pretty impolite to be LARPing in a busy train station}. 
The remaining comments expressed more straightforward emotions towards this phenomenon, accusing the MR street players of being \dquote{a nuisance to everyone} (L27), and calling them  \dquote{complete idiots to play in public places} (L2), and \dquote{dorky} to use HMD on streets.
In addition, the aesthetics and form factors of the MR devices were also a point of concern, with one comment comparing the experience to strapping \dquote{a heavy, awkward thing to the front of my face} (L5).
These responses reflect explicit social disapproval of public immersive MR games. They underscore the challenges of overcoming social awkwardness and perceived impropriety. It is crucial to address these concerns to foster widespread adoption and acceptance of this emerging form of entertainment in public spaces.

Regarding the on-site participants, only one participant showed slight hesitation and discomfort when first introduced to the headset in public, who later accepted it and tried the game (P20). Other on-site participants did not express feelings of awkwardness or initiate this topic in interviews until the researchers asked about their feelings of social awkwardness in the interview.

\subsubsection{Establishing social norms (7/63)}
\label{theme:socialnorms}

While some users expressed concerns about the social awkwardness and perceived impropriety of using MR technology in public spaces, others held a more optimistic view. They believed that the adoption of new technologies has historically led to changes in social norms, and MR street play is likely to follow a similar path.

Several users drew comparisons to the normalization of other technologies and behaviors that were once considered socially unacceptable or embarrassing. One participant noted that talking with headphones is now socially acceptable, despite the initial awkwardness of \dquote{talking to the air} (P20). Another user pointed out that glasses, umbrellas, and bikes were all \dquote{embarrassing at first until it becomes normalized} (L23). Some participants expressed confidence that technological evolution would lead to reduced headset sizes, drawing parallels to the historical miniaturization of mobile phones from their initial \dquote{brick-like} dimensions to current compact forms that have gained widespread social acceptance (L12).

The widespread adoption of Pokémon GO was also cited as an example of how a seemingly \dquote{dorky} behavior can become socially acceptable when enough people engage in it. Users agreed that people would wear headsets outside if the experience is compelling enough, just as there was \dquote{a plague of arms-length phone-holders in every park for a year} during the height of Pokémon GO's popularity (L25). The social acceptability of the behavior was further reinforced when enjoyed in the company of others, as one user noted, \dquote{If you and a friend are doing this together, you're no longer a dork} (L31). These comments suggest that there are people who believe that as MR technology becomes more compelling and widely adopted, the social norms surrounding its use in public spaces are likely to evolve. 

However, some others pointed out that the social norm surrounding MR technology in public settings has yet to be established (4/63). One user highlighted the difference between the social acceptability of looking at a phone and engaging in MR street play, stating, \dquote{Waving a stick around and shouting fireball while wearing a goofy headset is, at present, still unacceptable} (L8). Others suggested that the widespread adoption would require the integration of a popular AR game like Pokémon GO (L11).

The form factor of the headset device was identified as a crucial factor in determining the social acceptability of MR street play. As one user pointed out, \dquote{This (HoloKit with iPhone) is (yet) not the form factor that will change the paradigm, the smartphone is. Whoever gets the AR glasses that look like glasses out first wins the game} (L14). Another user doubted the establishment of a new social norm surrounding MR in the reply, stating that \dquote{This will only happen if the devices start to become indistinguishable from a normal pair of glasses} (L12).

\subsubsection{Gradient of Social Awkwardness} 
\label{theme:gradient_awkwardness}

Our findings revealed that a gradient of social comfort levels emerged across different modes of engagement: multiplayer play felt the most natural and least awkward, while solo play introduced some discomfort. Watching others play without access to the spectator view was described as the most awkward experience.

First, players in multiplayer games are less likely to feel awkward compared to single-player experiences. They feel more comfortable when the MR world is shared with another person, knowing they are not alone in MR while being observable to passersby. From the passersby's perspective, watching the players' interactive body movements gives them hints about the content of the MR game. When questioned about awkward levels in public settings, all on-site participants reported a lower sense of awkward feelings during paired games than single-player games. P1-P12 agreed that two people playing the game out there is fully acceptable. One of them stated, \dquote{It's no longer embarrassing if it takes two doing embarrassing things together} (P12).

Secondly, participants are also less likely to feel embarrassed than spectators or passersby. When participants recalled their feelings before they tried the game, they described feeling confused and anxious, as they did not know what was happening in the MR experience. In contrast, after the game, even if playing in single-player mode and behaving abnormally, many claimed they did not feel embarrassed but were eager to explain the content and gameplay to others. As one participant put it, \dquote{It's not embarrassing to play, once you start it's someone else's concern to be embarrassed} (P6).

% However, opinions vary on whether playing with strangers is more awkward than playing alone in public spaces. The spectator view can help alleviate awkwardness, but does not fully change people's willingness. In general, more introverted individuals tend to stick to close friends, while some extroverted people prefer playing with others over playing alone in public. 

% One introvert participant explained, \emph{"If you are a stranger and can't see spectator views, you may feel silly; If you are friends, you can anyway interact regardless of the spectator view" } (P8).

\subsubsection{Safety and security concerns (7/63)}
\label{theme:safety}

Many people expressed practical concerns about the potential safety issues that immersive MR street play may result in (5/63). There is a common concern about collision accidents due to the players being too immersed in the MR game and ignoring the real-world context among online respondents. One person commented about the MOFA videos, \dquote{I do worry that one of my kids is going to get trampled by an adult who can't see because he's too busy slaying a dragon} (L17).
Another major concern is about the HMD device being stolen (2/63). People worried that putting an iPhone in the headset would bring a \dquote{significant risk of phone theft in major cities} (L24).

\subsubsection{Divergence of online and on-site attitudes}
\label{theme:divergence}

Our analysis showed that the sentiment of online comments was more negative than that of on-site participants.

Moreover, the content of negative feedback also differs. On-site negative comments are often about specific technical limitations, including anchoring issues, the field-of-view limitations, and tracking latency (11/32, see 5.3.4). These critiques are typically framed in a manner that suggests a desire for improvement, with a more subdued and neutral tone. Most on-site players (31/32) did not express concerns about social acceptance until the researchers prompted them on this topic.

In contrast, online feedback is significantly more critical and confrontational. Online comments frequently raise pointed questions challenging the social acceptance of the game (13/31), potential safety risks associated with gameplay (3/31), MR headset form factors (3/31), and the overall entertainment value of the content (2/31).

\subsection{Gameplay Experience}
\label{theme:gameplay}

On-site participants generally described their game experiences as \dquote{fun}, \dquote{amazing}, \dquote{incredible}, and \dquote{magical} (31/63). They particularly appreciated that the game was intuitive to understand and operate, and that they could play and communicate with other players. However, some players also expressed frustration due to technological limitations, such as heavy headsets that caused discomfort on their noses (5/63).

\subsubsection{Fantasy world overlaid on the real world (23/63)}
\label{theme:fantasy}

Participants on-site reported that IMRSP provided them with a unique opportunity to immerse themselves in a magical fantasy world that seemingly came to life in reality. Players expressed a strong sense of being transported into the realm of their dreams and fantasies, where they could embody the roles of powerful wizards and engage in long-awaited duels. One player (P13) exclaimed, \dquote{I have been waiting for this duel all my life!}  Another participant expressed her excitement: \dquote{You have fulfilled my wildest fantasies. This is why I'm here} (P14).
    
The game evoked a strong emotional connection to fantasy lovers and their childhood dreams. A participant (P25) commented: \dquote{I feel like a wizard. I think my letter from Hogwarts is finally gonna come in the mail after all these years.} Referencing the Harry Potter series, their words showcased how this game tapped into the collective imagination and allowed players to embody the roles of their favorite fictional characters. These reviews show that the seamless integration of immersive MR technology and fun gameplay contributed to a deep sense of immersion, making the fantasy feel tangible and real for the players.

% For this, the game was also perceived as creative and pioneering in MR design, as mentioned by one art student \emph{"I have never know mixed reality can be played in this way"} (P9), and be commented as \emph{"promising"} (L14, P22).

\subsubsection{Wand as a mediator for intuitive interaction and narrative immersion (33/65, behavior)}
\label{theme:wand}

On-site participants praised the intuitive and smooth interaction facilitated by the combination of the wand and Apple Watch in this game. Unlike traditional handheld controllers, this design allowed for natural and immersive casting of spells in the real world. Participants quickly grasped the gameplay mechanics. In the on-site study, 24/65 participants were observed to immediately understand how to cast spells after the game started, either by listening to oral instructions or observing other players. Another 31/65 participants learned the mechanics in less than one minute with guidance from organizers and fellow or previous players. As one participant (P15) remarked, \dquote{It's so easy to use (the wand) too, very intuitive and functional.} 

The wand also served as a crucial element in the narrative design, providing a compelling explanation for the asymmetrical perspectives between players and passersby. By assuming the role of wizards and perceiving passersby as muggles, players could engage in the MR experience smoothly. The physical presence of the wand helped bystanders understand the ongoing MR experience, even without an MR view. \dquote{It makes much more sense to see people waving a wand. This is far better than randomly punching the air with your hand. } (P4).

The significance of the wand was further highlighted during a demo where it was taken away from a player's hand in the later stages of the game. The player immediately struggled to perform as she had been doing just 5 seconds earlier, expressing her confusion and requesting the wand back: \dquote{Now I forgot how to move… Can you give me back the wand?} (P8).
This incident demonstrated how the presence of a prop can support players' understanding and engagement and maintain the narrative immersion.

\subsubsection{Exercise and workout (14/65)}
\label{theme:exercise}

The game led to a notable increase in physical activity and exercise among players. Many participants (26/65, behavior) were observed breathing heavily after playing, indicating significant physical exertion. The level of physical activity varied depending on the game mode, with the single-player game resulting in minimal movement, the duel game encouraging more movement and dodging with the opponent, and the dragon game resulting in the most physical activity, with players running to chase the dragon or avoid fireball attacks.

Participants appreciated the physical demands of the game, with some expressing surprise at the level of exertion required: \dquote{This is such a workout!} (P16, P17). Another compared it to playing a \dquote{fierce badminton game} (P9), suggesting that MOFA could serve as an engaging alternative to traditional forms of exercise. The physical nature of the gameplay was also seen as a distinguishing factor compared to traditional PC games, as P7 noted: \dquote{The main diﬀerence from PC games is its physical exertion.}

\subsubsection{Frustration caused by technological inadequacies (12/65)}
\label{theme:frustration}

Despite the overall positive gameplay experience, some participants expressed dissatisfaction with certain technological aspects of MOFA. 

Participants sometimes raised issues about the precision of the anchoring system. \dquote{Sometimes the shield is here and there} (P3). Some other participants found the user flow of the co-location connection to be cumbersome. For example, \dquote{I think the connecting could be a bit easier} (P18).

Participants also noted limitations of the current MR display technology, such as the transparency of the optical see-through AR and the limited field of view. For example, \dquote{it can only show a part of the dragon, and I feel that it's trapped in the screen} (P11).

Participants also complained about the lack of prompt feedback to their body movements. \dquote{It didn't look like I sent it out with my own hands. The angle of my hand, the angle of my body, and the angle of my head will not affect the result} (P7). Similarly, another participant (P8) stated, \dquote{I have no way of judging how precise I am going to be, it's hard to aim.}

\subsection{Environment and Location}

    % real word as game terrain - behavior
    % background impact
    % environment and social acceptance
    
The environment provides context for gameplay, influencing excitement, movement, and the degree of social acceptance.

\subsubsection{Landscape acts as a vital backdrop}
\label{theme:landscape}

The surrounding environment setting can be seen as the backdrop of the game map and plays a crucial role in players’ emotional engagement. A proper backdrop of landscape allows players to fully engage with the game's mission and narrative in a meaningful way. Environments that fit the game content can enhance the game experience. For example, researchers once organized to demonstrate and play MOFA ``the Dragon'' in an alpine mountain. The natural environment and stunning scenery evoked a sense of awe in participants through the epic scale. The small group of players and spectators started to call the players \dquote{heroes} and \dquote{heroines} in the process. After the test, one of them commented that \dquote{Today I feel like I am in a dream} (P30).
On the other hand, the same game experience can be diminished in environments that do not align with the thematic elements of the game. For instance, attempting to play the dragon-hunting game in a small, low-ceiling interior setting may feel less immersive and impactful.

\subsubsection{The real world acts as the digital game terrain (8/65)}
\label{theme:realworld}

% Context-aware movement of both the digital world and the physical world
The design of MR environments is not solely confined to the digital realm; the physical environment serves as a \dquote{game terrain} that influences players' behaviors in games. In ``The Duel'', participants frequently navigate around pillars or tables for cover and shielding (P3, P4, P11, P12). The MOFA games did not integrate real-time environment detection in gameplay, meaning that obstacles in the physical world do not impede attacks; moreover, spells and shields are not geometrically occluded in the rendering, yet players still use real objects for cover. Nevertheless, players still inherently engage in these strategies, as their vision tracks opponents' movements rather than merely relying on MR content. Participants also expressed expectations of future gameplays reacting to the physical environment, like spells bouncing off walls and grounds, or \dquote{imagine a Charizard (Pokémon GO dragon) standing on top of my local mall} (L22), considering \dquote{The connectivity to the building and the space will be the strong point of the game} (P12).

\subsubsection{Context shapes social acceptance}
\label{theme:context}

The spatial context, including the environment and location, plays a crucial role in determining social acceptance and adherence to social norms when it comes to immersive MR street play. People perceived that open spaces, such as parks, are more suitable, while crowded and busy spaces or potentially unsafe situations, like sidewalks, are considered inappropriate. One online reviewer (L13) expressed a strong objection to the video depicting people playing the MOFA games: \dquote{The park, yes, that works, but a busy street, are they just out of touch with reality?} 

People who favored the game envisioned that the success of the game should be tied to open areas in urban parks rather than crowded areas. For example, L16 commented that  \dquote{This is doing all the right things and I think I'll get one (the game), though I think their videos are doing a disservice showing people playing in crowded areas. Show people playing in parks surrounded by people bored on picnic blankets...in Dolores Park when the weather turns nice in April.} 
Certain venues also provide social acceptance to the immersive MR street play, as L15 commented \dquote{If Pokémon GO had this during its heyday, the dork factor would have been socially acceptable.}
 % elan
\section{Discussion}
\label{sec:discussion}

\subsection{Immersive MR Street Play in Public: From Dream to Reality}

Visions of MR headsets transforming public street play have been widely disseminated across various media, from films and promotional tech demonstrations to game trailers. While this remains a ``dream'' for tech enthusiasts, most people have not yet experienced this kind of play. Through realizing this dream of deploying our open-source exploratory game probe series MOFA into the real world, we glimpse the future of IMRSP, providing an initial yet foundational understanding of the social implications of this novel form of play.

\subsubsection{Experiencing IMRSP In the Wild}

Although prior HCI research has extensively explored interactive public systems---such as large public displays \cite{Cao2008Flashlight}, urban AR games such as Pokémon GO \cite{Paavilainen2017Pokemon}, human-robot interactions in city settings \cite{Pelikan2024Encountering}, and VR remote collaboration scenarios within constrained passenger spaces \cite{Medeiros2024Exploring,Ng2021Passenger,Medeiros2023Surveying}—the specific context of co-located, bodily mixed reality play using see-through MR headsets in actual urban streets remains underexplored. Our study directly addresses this gap by providing one of the first empirical investigations of IMRSP within authentic street environments, beyond controlled or private settings.

\begin{quote}
\dquote{A dream you dream alone is only a dream. A dream you dream together is reality.}  \flushright --- Yoko Ono
\end{quote}

Our in-the-wild approach revealed highly positive reactions from participants. Players consistently described IMRSP as immersive, engaging, and significantly more enjoyable than initially expected. A crucial aspect highlighted by players was the seamless integration of physical movement, digital interaction, and social connectedness. Unlike VR, often perceived as isolating, participants appreciated how IMRSP facilitated shared, collaborative experiences with friends in real-world surroundings. These findings underscore that MR street play offers a viable, compelling alternative to solitary screen-based gaming, reinforcing its potential as an inclusive and socially enriching public activity.

Furthermore, our observations demonstrate how IMRSP actively reshapes everyday urban social dynamics, not merely overlaying digital content onto physical spaces but transforming streets into interactive stages for playful encounters. Echoing earlier research that identifies the street as a vibrant, inhabited site of socially organized human action \cite{Pelikan2024Encountering, Stevens2007Ludic}, our study found that MR street play creates shared, performative experiences among players, bystanders, and passersby. The urban street environment---traditionally governed by subtle social norms and unspoken cues—becomes a dynamic arena of spectacle and interaction, actively drawing passersby into novel social engagements.

\emph{Our empirical findings suggest that widespread adoption of IMRSP could lead to significant shifts in social norms, behaviors, and the design and use of urban public spaces.} This study thus begins to unpack what might happen if MR street play transitions from niche demonstrations to mainstream urban activity, highlighting the necessity to understand and carefully manage the interplay between technological innovation, social dynamics, and spatial design in public environments.

\subsubsection{Augmenting Social Dynamics through IMRSP}

Our findings indicate that IMRSP significantly reshapes social dynamics in public spaces, aligning closely with established concepts from prior research on public interactive installations. One notable parallel is the ``honeypot effect'' \cite{Wouters2016Uncovering}, previously observed in interactive displays and street-level activities \cite{Cao2008Flashlight, Reeves2005Designing}, where visible, playful activities naturally draw passersby into becoming spectators and even spontaneous participants. Similarly, during our field trials, once a small group initiated MR street play, curious bystanders consistently gathered around them. Players immersed in digital battles or performing dramatic gestures transformed ordinary streets into lively theaters of interaction. This mirrors previous findings showing how street play and public interactive technologies dynamically convert urban spaces into stages for public performance and social interaction \cite{Stevens2007Ludic, Benford2011Performing}.

Beyond merely drawing attention, IMRSP created significant social contagion effects. Our observations demonstrated that players felt notably less awkward engaging in playful activities when accompanied by peers. The presence of a supportive social group reduced inhibitions, echoing findings from previous interactive installations where collective engagement validated individual participation in otherwise unconventional behaviors \cite{Reeves2005Designing, Wouters2016Uncovering}. Additionally, bystanders frequently mirrored the players’ enthusiasm through gestures, applause, or informal engagement, reinforcing the playful atmosphere. This contagion effect underscores IMRSP’s potential to leverage social facilitation, effectively lowering barriers for hesitant individuals to engage in public MR interactions.

Moreover, participants frequently embraced the public attention as an opportunity for deliberate showmanship \cite{Reeves2005Designing}, exaggerating movements and reactions to entertain and involve bystanders. Rather than feeling inhibited by an audience, these players became energized, further emphasizing the performative and expressive dimensions of IMRSP. This aligns with prior research on embodied interaction and performance-led HCI approaches, where technology encourages expressive and performative engagement \cite{dourish2001action, Benford2013Performance}. Such intentional performances amplified community interactions and fostered a positive feedback loop, attracting additional spectators and participants.

\emph{Collectively, these findings reveal how IMRSP actively transforms public spaces into shared arenas of playful social interaction and performance.} The street, traditionally governed by subtle social norms, becomes a vibrant social stage where traditional boundaries between player and audience blur. As MR street play grows more prevalent, understanding and managing these emergent social dynamics will be essential for facilitating inclusive and positive public interactions in MR-rich urban futures.

\subsection{The Frictional Path to Widespread MR Street Play Adoption}

\subsubsection{Contrast of Social Acceptance: On-site vs. Online Perspectives}

Our study revealed a notable gap in social acceptance between those who experienced IMRSP firsthand and those who observed it remotely through recorded gameplay videos. Participants who directly engaged in MR street play consistently reported highly positive experiences, describing IMRSP as enjoyable, immersive, and socially rewarding. They appreciated the embodied interaction facilitated by MR headsets, highlighting that playing alongside friends and maintaining awareness of their physical surroundings significantly reduced feelings of isolation or awkwardness. This on-site enthusiasm aligns closely with prior research emphasizing the role of collaborative experiences in alleviating social discomfort during public interactions \cite{Reeves2005Designing, Huggard2013Musical,Benford2013Performance}. 

In sharp contrast, remote online observers, who lacked direct interaction and immersion, expressed far more ambivalence and skepticism towards IMRSP. Online commenters frequently voiced concerns about social awkwardness or embarrassment associated with observed behaviors, often characterizing public MR play as ``strange'' or ``inappropriate.'' Safety concerns were also prominent among these remote viewers, who worried that players might become distracted, increasing the risk of accidents or collisions in public settings. These critical perceptions resonate with earlier societal reactions to novel wearable technologies such as Google Glass, where non-users perceived the technology as intrusive or socially disruptive due to unfamiliarity and lack of direct exposure \cite{Due2015social, Denning2014situ, Lebeck2018Security}.

This perception gap underscores a critical challenge in achieving broader public acceptance of IMRSP. On-site players, immersed in the embodied and collaborative MR experience, quickly grasp its social and recreational value, seeing the technology as socially acceptable and enriching. Conversely, external observers, lacking firsthand immersion, primarily witness unconventional behaviors without the contextual understanding or immediate social engagement afforded by direct participation. As such, these observers often interpret IMRSP as violating established social norms or expectations, particularly regarding appropriate behavior in shared public spaces---a reaction aligned with Goffman's insights on social norms and the management of impressions in public contexts \cite{goffman2008behavior}.

\textit{Our findings thus highlight that firsthand immersion and collaborative participation significantly influence social acceptance and positive evaluations of MR street play.} To bridge this acceptance gap, future design and deployment strategies might emphasize facilitating bystander comprehension of MR activities or provide accessible avenues for casual participation. Promoting greater public familiarity and understanding of MR technology can mitigate initial resistance and skepticism, paving the way for more inclusive and socially integrated MR experiences in public spaces.

\subsubsection{Ethical, Safety, and Privacy Concerns in Public MR Play}

Despite the positive social engagement observed during IMRSP experiences, our findings indicate significant ethical, safety, and privacy concerns 
\cite{Katins2024Assessing,Millard2024Ethics} that must be addressed to facilitate broader social acceptance \cite{Eghbali2019Social}. A prominent issue is the asymmetric power between MR headset wearers and non-participating bystanders. MR users have privileged access to hidden digital information through outward-facing sensors, such as cameras and depth scanners, potentially recording bystanders without explicit consent or awareness. This dynamic creates a sense of vulnerability among non-users, who might perceive MR technology as intrusive or threatening their privacy—an effect reminiscent of earlier public reactions to technologies like Google Glass, where similar privacy concerns led to significant public backlash \cite{Bajorunaite2024VR}. Recent research also underscores the necessity for privacy-enhancing mechanisms that account for bystanders’ varying preferences for awareness and consent in everyday AR scenarios \cite{OHagan2023Augmenting}.

Physical safety represents another critical concern identified in our research. Although no incidents occurred during our field studies, the immersive, fast-paced nature of IMRSP raises genuine risks of accidents or collisions. Players engrossed in virtual gameplay might inadvertently overlook physical hazards, such as vehicles, street furniture, or pedestrians. Online observers particularly highlighted these potential dangers, reflecting valid concerns about player distraction in dynamic public environments. Additionally, social safety---how comfortable and secure bystanders feel around MR participants---must be carefully considered. Sudden or unpredictable movements by MR players could alarm or unsettle individuals who are unfamiliar with the activity, creating unintended social tension. As social MR systems grow more pervasive, the adoption and effectiveness of in-situ safety tools to mitigate harassment, discomfort, or unwanted encounters become increasingly crucial for ensuring both player and bystander well-being in shared environments \cite{Weerasinghe2025Mute}.

Furthermore, IMRSP explicitly challenges established social norms, particularly Goffman's concept of ``civil inattention,'' where individuals typically ignore each other in public settings to maintain social order \cite{goffman2008behavior}. By introducing visible, unconventional behaviors into shared public spaces, MR street play inherently disrupts these existing norms, compelling bystanders to acknowledge and respond to unusual public performances. While some individuals might appreciate these playful disruptions as novel or refreshing, others could interpret them negatively, perceiving such activities as intrusive or socially inappropriate. This tension could potentially lead to social friction or resistance against MR enthusiasts, highlighting the need for proactive design and policy considerations to manage public interactions sensitively. Additional scholarship highlights that the widespread adoption of everyday AR will introduce significant ethical challenges---such as manipulation and information disorder---necessitating clear guidelines for who can mediate reality and for what purpose \cite{OHagan2023Augmenting}. Building on this, recent work argues that everyday, pervasive AR demands a new societal consensus on ``perceptual rights'' to protect both users and bystanders from potential misuse and abuse \cite{OHagan2024Viewpoint}.

\emph{Addressing the social and ethical challenges in IMRSP requires careful attention to transparency and consent regarding MR technology usage, ensuring clear communication about data capture and privacy implications.} Emphasizing group-oriented rather than solitary gameplay may reduce perceptions of awkwardness or intrusiveness~\cite{Deterding2018Alibis}. Moreover, urban spatial planning that clearly defines appropriate public spaces or dedicated zones for MR activities could enhance both physical and social safety. Proactive measures addressing privacy, safety, and public etiquette concerns will be essential for the successful integration and acceptance of IMRSP in everyday urban life, preventing potential social backlash and fostering a harmonious coexistence between MR players and the broader public.

\subsection{Methodological Implications for MR Futuring}

In this paper, we use today's MR technology to enact MR's own futures. Our methodology employs a fully-functional high-fidelity research probe that instantiates a plausible next-step MR experience \emph{in situ}. This enables observation of real social consequences---bystander dynamics, on-the-fly norm negotiation, and emergent coordination---rather than relying on hypothetical narration or studio vignettes. Methodologically, we blend research-through-design \cite{Zimmerman2007Research} with immersive speculative enactment \cite{Simeone2022Immersive}. However, our approach differs from prior immersive speculative enactment works that typically use VR to depict futures from within a headset. Instead, we use today's MR itself to probe MR's own future, deployed in public space with real passersby. This yields ecologically valid evidence about how people move, signal, and coordinate around MR systems—evidence that is hard to surface in purely fictional or studio-bound enactments.

This approach is effective because, although MR hardware will plausibly improve (e.g., smaller form factors, wider FOVs, better tracking), the core social-cognitive frictions that emerge in public, multi-party settings are likely to persist. MR technology-mediated experiences obscure what others are attending to, hindering theory of mind \cite{McKenna2023Theorya} and making norms less visually legible to bystanders who cannot see what the wearer sees \cite{VonTerzi2021TechnologyMediated,DesnoyersStewart2025Being,Hagan2023Privacy}. Passthrough and layered visualizations further complicate mutual understanding in public places \cite{Bailenson2024Seeing}. Additionally, MR renders reality programmable: MR designers effectively codify ``strange new rules'' \cite{rao2025strange}---what we call ``(mixed) reality protocols'' \cite{Hu2025Autonomous}---that determine what counts as visible, actionable, or socially appropriate within an app. Because these protocols vary by app, they can drive divergent behaviors in the same physical place, further hindering theory of mind even between MR wearers. This limitation will likely persist even as hardware improves. The embodied signatures we observe---unshared referents, ambiguous gestures, and micro-breaches of street etiquette---are system-agnostic. They emerge from asymmetries of technology-mediated perception and accountability rather than from any single device's limitations. Consequently, findings from today's deployments speak to the long arc of MR sociality: as people encounter device-specific, app-specific rules that aren't universally visible, they will continue to negotiate trust, accountability, and shared context in public.

Fortunately, thanks to MR's programmability, many of tomorrow's rules can be implemented today: we can encode prospective interaction logics, object behaviors, and social norms on current hardware; deploy them in real environments; and study how trust, accountability, and shared context are actually negotiated. In essence, MR offers a prefigurative testbed: by programming futures into the present, we can derive empirically grounded implications for those futures, echoing William Gibson's famous insight: 
\begin{quote}
\dquote{The future is already here---it's just not very evenly distributed.} \flushright --- William Gibson
\end{quote}

Methodologically, this paper suggests three commitments for MR futuring. First, favor enactment over narration: stage near-future MR experiences in public settings where bystander interpretation, coordination, and etiquette are genuinely at stake. Second, treat MR prototypes as \emph{norm engines}: vary the reality protocol (what is visible/doable/appropriate) and observe how rule changes reconfigure gesture legibility, referent alignment, and accountability. Third, analyze what persists: focus on social frictions rooted in asymmetric perception and theory-of-mind constraints, which are likely to endure across device generations. 

In sum, by implementing speculative rule sets now and examining their social consequences in real-world settings, our method provides a concrete pathway for researchers and designers to co-produce MR futures with the public. This approach goes beyond merely imagining possibilities---it allows us to observe how these futures actually unfold in practice. We anticipate this methodology will inspire additional work that engages users in prospective MR frameworks, thereby advancing both design practice and our empirical understanding of emerging social norms.

\subsection{Design Recommendations for Socially Acceptable IMRSP}

Based on our game development experience, findings, and the above discussion, we distill several key design suggestions for future IMRSP development.

\subsubsection{Designing for Contextual Awareness}
Given its co-located interactions in public spaces, we propose that IMRSP designs should explicitly consider contextual awareness \cite{Grubert2017Pervasive} across multiple dimensions. First, designers should critically evaluate the social norms and cultural acceptability of proposed embodied interactions within specific public contexts. This includes designing game themes appropriate for these contexts. Second, embracing the multiplayer aspect of IMRSP, mechanics that encourage collaborative engagement not only enhance the play experience but also mitigate social awkwardness through collective participation. Third, designs should incorporate adaptable interaction thresholds that allow users to modulate their performative engagement based on environmental factors and personal comfort levels. Fourth, since IMRSP transforms public spaces into performative arenas, designers should intentionally craft spectator experiences \cite{Reeves2005Designing} that facilitate understanding and potential participation. Finally, we recommend that designers systematically analyze contextual parameters---including spatial characteristics, cultural specificity, and social dynamics---before designing IMRSP applications in public settings.

\subsubsection{Harmonizing with Social Norms through Explainable Social Affordances}
As MR street play disrupts conventional social norms, it creates a disconnect between players' social and virtual identities, potentially causing social awkwardness. There is a need to design MR street play that promotes social etiquette and manages instances of civil inattention \cite{goffman2008behavior} in public settings. This can be achieved by incorporating narrative elements, role design, and physical props that signal ongoing MR activities to bystanders, thus reducing confusion and promoting mutual awareness. For example, in our MOFA game, we emphasized the use of physical props, such as wands, to rationalize players' behavior. Given the narrative roles of wizards and witches, these props bridge the gap by offering a tangible connection to the player's in-game persona, enhancing player immersion while alleviating bystanders' confusion. These social affordances~\cite{Isbister2018Social,Wu2020Megereality} also help lower the learning curve for players. Furthermore, incorporating additional elements like music, costumes, and visual cues can enrich the gaming experience and enable clearer communication of player behaviors to the public. 
Finally, designers can help users navigate social boundaries by providing guidelines on appropriate behaviors within different social and cultural contexts. This might include features that notify users when they enter culturally sensitive areas or crowded spaces where MR play could potentially cause social friction.

\subsubsection{Designing Engaging Movement while Maintaining Safety}

IMRSP naturally engages users' movement as its interaction mechanism. We recommend four design considerations to ensure both engaging gameplay and public safety. First, maintain a largely unobstructed, transparent real-world view in the MR interface to help players remain aware of their physical surroundings, thereby reducing collision risks when moving in public spaces. For example, in MOFA, augmented overlays are intentionally minimal to prevent users from losing their sense of reality: most of the display area in MR should remain free of UI elements and augmented content. Second, design interaction mechanics that limit potentially dangerous movements---for example, MOFA's magic shield system provides clear attack targets without requiring physical contact between players. Third, incorporate visual feedback systems that substitute for physical impacts or UI elements. For example, the shield-breaking animations signify instant feedback of attack, avoiding the need to overlay traditional health bars over opponents. This creates visceral feedback without requiring players to frequently check UI elements for status updates, which would distract attention and potentially cause fatigue. Finally, include spatial boundary indicators that dynamically adjust based on surrounding pedestrian density, helping players maintain appropriate play zones in varying public conditions. These design principles help balance the inherently movement-based nature of IMRSP with the safety considerations necessary for deployment in shared public environments.

\subsubsection{Guiding Users to Appropriate Gameplay Locations}
Respecting the importance of contextual flexibility and spatial appropriateness in urban design, MR street play should guide users to open, adaptable spaces. Parks, pedestrian zones, and urban plazas are ideal environments for such activities, offering ample room for movement while minimizing the potential for physical or social collisions. Urban spaces can benefit from design features that subtly guide MR use. For example, visual markers or delineated zones can indicate areas where MR activities are encouraged, allowing participants and bystanders to understand the social expectations within these spaces.

\subsection{Limitations and Future Work}
This paper presents an initial exploration of IMRSP through our game probe MOFA, demonstrating the capabilities of current affordable XR devices to support MR street play in the wild. While this approach yielded valuable insights, we acknowledge several limitations and propose a research agenda for future work on IMRSP.

\subsubsection{In‑the‑Wild Scenarios} 
In this study, we examined only five scenarios where IMRSP was deployed, which may not fully represent everyday ``in-the-wild'' conditions. For instance, in exhibition contexts such as expos and conferences, participants were predisposed to encounter novel technologies, creating an atmosphere of heightened tolerance for unconventional embodied interactions. Similarly, campus settings, where physical activities are commonplace (e.g., frisbee games), may have provided a more accommodating social context than truly public spaces. These settings likely attracted participants with greater openness toward experimental gameplay, potentially mitigating social acceptability concerns that might arise in genuinely public environments. This contextual bias may partially explain the divergent perspectives between in-person and online commenters, where online respondents emphasized the potentially disruptive aspects of the game, while in-person participants primarily focused on positive elements, experiencing minimal social awkwardness as others actively engaged in the experience. As this research represents an initial exploration of IMRSP, future studies should extend to more ecologically valid scenarios similar to how Pokémon GO operates across diverse public environments, to better understand the social dynamics and acceptance of such technologies in authentic everyday settings. Future work could look into the following directions:

\paragraph{Explore social acceptance factors for IMRSP} Future researchers can explore longitudinal research on social acceptance factors blocking the widespread adoption of IMRSP. It would be meaningful to understand why people initially reject this game online but accept it after experiencing it in person. They can even invite those who reject the game online to try it on-site, allowing them to study user reactions and long-term retention behaviors. This research would help identify the key barriers preventing this type of game from becoming successful. 

\paragraph{Facilitate intent inclusivity in public spaces for spectatorship} As described in \cite{Hu2024Intent}, facilitating intent inclusivity includes understanding the intent differentiation of HMD wearers and non-HMD wearers. For HMD wearers, this may include unconcerned players, privacy-conscious players, acceptable non-sharers, eager sharers, and casual inviters. For non-HMD wearers, this may include enthusiastic recruiters, indifferent passersby, tolerant observers, disturbed onlookers, curious spectators, included spectators, and those ready to join with handheld AR or head-mounted displays. Studying these different intents could lead to more inclusive and socially aware IMRSP experiences.

\subsubsection{Gameplay} 
Our study's MR street play experiences were based on a single game probe---MOFA, which features three games within a magic-themed context. While this theme provided a shared framework for players to quickly understand gameplay mechanics, it may not fully capture the diverse range of potential MR street play scenarios. Although we tested only one magical fantasy theme, we believe this approach can be applied more broadly. Future studies should expand beyond magic and explore games with different themes, interaction styles, and levels of physical engagement to better understand the full spectrum of MR street play possibilities and their effects on social dynamics. We outline two near‑term steps: 

\paragraph{Explore the design space for IMRSP} This field remains largely uncharted, so we encourage researchers to investigate its various aspects and potential. This includes exploring different game mechanics, interaction modalities, and social dynamics unique to IMRSP.

\paragraph{Identify effective social affordances that communicate social norms in IMRSP} Similar to the wands used in MOFA, future work should design and test various physical props, visual cues, and auditory signals that help convey players' actions and intentions to bystanders.

\subsubsection{Physical Interaction} Due to the game design and safety considerations, all three games in MOFA required no physical contact, which limited our exploration of potential physical interaction dynamics in MR street play. We acknowledge the potential of physical contact in introducing richer, more complex social interactions, which might enhance or complicate MR experiences in public. Future studies might explore MR street play scenarios that permit safe, controlled physical interactions to better understand how touch and close proximity impact social bonding, comfort, and behavioral dynamics in public MR environments. In response to this limitation, future work should: 

\paragraph{Develop ethical, privacy, and safety guidelines for IMRSP} These guidelines should address issues such as player privacy, physical safety in public spaces, and the potential psychological impacts of immersive experiences in urban environments. In particular, developing protocols for inviting bystanders to join IMRSP is essential. What permissions are necessary? How can we protect privacy? Which invitation methods are most effective? This research could involve testing various connection strategies and analyzing their effectiveness and social acceptability.

\subsubsection{Environmental Interaction}
The environmental interaction in our study was limited, as the MOFA games were designed primarily to encourage social interaction among players rather than interaction with the surrounding physical environment. While players could move freely within certain spaces, the games did not incorporate elements of the physical environment, such as urban infrastructure, landmarks, or other spatial features, which could enhance immersion and contextual relevance. Future studies could explore MR street play designs that integrate environmental interactions, such as using real-world objects as interactive components or adapting gameplay based on specific environmental characteristics.

% \subsection{Future Research Agenda}
% For future research on IMRSP, we propose the following agenda:

% \subsubsection{Explore the relationship between Live Action Role Play (LARP) and IMRSP} How can mixed reality enhance LARP \cite{Johansson2024Why} experiences? This research could involve adapting existing LARP scenarios to IMRSP contexts and analyzing the benefits and challenges of this integration.

\paragraph{Theoretical exploration of IMRSP's relationship to adjacent forms of play and games} For future research on IMRSP, we suggest a theoretical exploration of its relationships with various fields, such as social games \cite{Goncalves2023Social}, urban games \cite{DeLuca2012Urban,Woolley2013Exploringa}, alternate reality games \cite{Kim2008Alternate}, playful city \cite{Nijholt2017Playable}, playful wearables \cite{Buruk2023Playful}, pervasive games \cite{Ahn2016Pervasive}, location-based games \cite{Laato2019Review,Sutko2011Locationaware}, mixed reality games \cite{Bonsignore2012Mixed,Millard2024Ethics}, live action role play (LARP) \cite{Johansson2024Why,MarquezSegura2018Designing}, sport games \cite{Mueller2011Designing}, and exertion games \cite{Mueller2008Taxonomy,Mueller2016Exertion}. This includes investigating how IMRSP aligns with existing game theories and urban play concepts, examining its potential to transform urban spaces into interactive playgrounds, exploring its use for social change or education in urban environments, and analyzing its intersection with immersive technologies and physical activity-based gaming. 
 %rem 
\section{Conclusion}
\label{sec:conclusion}

This work presents our in-the-wild empirical studies of immersive mixed reality street play (IMRSP) using our open-source exploratory game probe series MOFA. Our findings demonstrate IMRSP's potential as a new paradigm for highly social and engaging exertion-based play in public spaces, supported by observations of rich social phenomena and dynamics. By comparing online reviews with on-site gameplay experiences, we identified key challenges in social acceptance and environmental factors affecting adoption. These insights inform our suggested next steps and research agenda for IMRSP. We conclude that, despite its engaging social nature, widespread acceptance of IMRSP may take longer than anticipated due to social, ethical, safety, and privacy concerns.

% We concluded that despite IMRSP novelty, the widespread acceptance of IMRSP in public spaces may take longer than anticipated. While some argue that, like smartphones, it will eventually change social norms, the extensive body movements required by IMRSP present a significant hurdle. Perhaps, like other sports, we'll eventually see dedicated playgrounds for IMRSP, similar to baseball or basketball fields, rather than it being played everywhere in the streets as promoted in those visionary videos.

% \section{Acknowledgement}
% Author Contributions:

% Author 1: Project lead, concept designer, game designer, and programmer. Conducted parts of empirical studies in the wild.

% Author 2: Responsible for literature review and theoretical discussion on empirical results.

% Author 3: User experience designer, illustrator. Led empirical studies, including in-the-wild research, on-desktop research, and thematic analysis.

% Author 4: Contributed to research concept discussion and writing.

% Author 5: Provided research guidance and writing support.

% As non-native English speakers, we employed an AI-based language tool to refine our manuscript's grammar and enhance its overall flow. %rem

\begin{acks}
This research was funded by Holo Interactive. We thank Yuchen Zhang and Sizheng Hao from Holo Interactive for developing the MOFA system.
We extend our gratitude to Professor Steve Benford and Professor Gisela Reyes-Cruz from Mixed Reality Lab at University of Nottingham for their valuable advice and suggestions. We especially thank Eva Yutzu Chen for assisting with user testing. We are grateful to students from the School of Design and Innovation at China Academy of Art, IDM and ITP students from New York University, and 706 Youth Community in Dali, China for supporting our user tests. Additionally, we appreciate the participants from SIGGRAPH, CHI, CHI Play, and UbiComp/ISWC onsite exhibitions for their valuable feedback that helped improve the system. 
\end{acks}

\clearpage
\bibliographystyle{ACM-Reference-Format}
\bibliography{reference}

\appendix
\newpage
\section{Semi-structured Interview Questions for Contextual Inquiry}
\label{appendix:interview}
\subsection{Immediate Post-play Reactions}
\begin{itemize}
  \item \textbf{First Impression:} What was your first impression right after playing? How do you feel now?
  \item \textbf{Excitement:} Which part of the game excited you the most?
  \item \textbf{Opponents \& Allies:} How did it feel to see your opponent or teammate casting spells?
  \item \textbf{Physical Activity:} Did it feel like you got some exercise? Was that fun, tiring, or both?
  \item \textbf{Interaction:} Did you find the spell-casting intuitive? Was the goal easy to understand?
  \item \textbf{Least Favorite Part:} What would you say is your least favorite part of the game?
\end{itemize}

\subsection{Social Dynamics and Concerns}
\begin{itemize}
  \item \textbf{Being Watched:} Did you notice any reactions from people around you? How did it feel to be observed by bystanders? 
  \item \textbf{Awkwardness:} Was there any moment you felt awkward or embarrassed?
  \item \textbf{Bystander Perspective:} (For bystanders) What made you want to stop and watch?
  \item \textbf{Comfort in Location:} Were you comfortable moving around the way you did here? Any concerns?
  \item \textbf{Safety and Concerns:} Did you feel safe playing in this spot? Do you have other concerns? 
  \item \textbf{Environmental Impact:} How did playing here (this park, plaza, etc.) affect your experience?
\end{itemize}

\subsection{Future Adoption}
\begin{itemize}
  \item \textbf{Previous Experience:} Have you played VR or MR before? Which do you prefer, and why?
  \item \textbf{Retention \& Recommendation:} Would you play this again? Under what conditions, or would you recommend it to friends?
  \item \textbf{Adoption Factors:} What could help or hinder you from playing more often in the future?
\end{itemize}

% \section{Technical Details of MOFA}
% \label{sec:tech}

\end{document}